\documentclass[twocolumn]{aastex631}

\newcommand{\Msol}{\text{M}_\odot}
\newcommand{\Mstar}{\text{M}_*}

\usepackage{graphicx}	
\usepackage{amsmath}	
\usepackage{amssymb}	
\usepackage{bm}		
\usepackage{orcidlink}

\usepackage[T1]{fontenc}
\usepackage{ae,aecompl}
\usepackage{newtxtext,newtxmath}
\usepackage{tablefootnote}

\shorttitle{Age Gradients of Dwarf Galaxies}
\shortauthors{Riggs et al.}

\begin{document}

\title{Testable predictions of outside-in age gradients in dwarf galaxies of all types}

\correspondingauthor{Claire Riggs}
\email{riggs@physics.rutgers.edu}

\author[0000-0001-8894-5671]{Claire L. Riggs}
\affiliation{Department of Physics and Astronomy, Rutgers, The State University of New Jersey, 136 Frelinghuysen Rd, Piscataway, NJ 08854, USA}

\author[0000-0002-0372-3736]{Alyson M. Brooks}
\affiliation{Department of Physics and Astronomy, Rutgers, The State University of New Jersey, 136 Frelinghuysen Rd, Piscataway, NJ 08854, USA}
\affiliation{Center for Computational Astrophysics, Flatiron Institute, 162 5th Avenue, New York, NY 10010, USA}

\author[0000-0002-9581-0297]{Ferah Munshi}
\affiliation{Department of Physics \& Astronomy, George Mason University, 4400 University Drive, MSN: 3F3, Fairfax, VA 22030-4444, USA}

\author[0000-0001-6779-3429]{Charlotte R. Christensen}
\affiliation{Physics Department, Grinnell College, 1116 Eighth Ave., Grinnell, IA 50112, USA}

\author[0000-0002-2970-7435]{Roger E. Cohen}
\affiliation{Department of Physics and Astronomy, Rutgers, The State University of New Jersey, 136 Frelinghuysen Rd, Piscataway, NJ 08854, USA}

\author[0000-0001-5510-2803]{Thomas R. Quinn}
\affiliation{Astronomy Department, University of Washington, Seattle, WA 98195, USA}

\author{James Wadsley}
\affiliation{Department of Physics and Astronomy, McMaster University, Hamilton, ON L8S 4M1}
\affiliation{Origins Institute, McMaster University, Hamilton, ON L8S 4M1}



\begin{abstract}

We use a sample of 73 simulated satellite and central dwarf galaxies spanning a stellar mass range of $10^{5.3}-10^{9.1}\Msol$ to investigate the origin of their stellar age gradients.  We find that dwarf galaxies often form their stars ``inside-out,'' i.e., the stars form at successively larger radii over time.  However, the oldest stars get reshuffled beyond the star forming radius by fluctuations in the gravitational potential well caused by stellar feedback (the same mechanisms that cause dwarfs to form dark matter cores). The result is that many dwarfs appear to have an ``outside-in'' age gradient at $z=0$, with younger stellar populations more centrally concentrated.  However, for the reshuffled galaxies with the most extended star formation, young stars can form out to the large radii to which the old stars have been reshuffled, erasing the age gradient. We find that major mergers do not play a significant role in setting the age gradients of dwarfs. We find similar age gradient trends in satellites and field dwarfs, suggesting environment plays only a minor role, if any. Finally, we find that the age gradient trends are imprinted on the galaxies at later times, suggesting that the stellar reshuffling dominates after the galaxies have formed 50\% of their stellar mass.  The later reshuffling is at odds with results from the {\sc fire-2} simulations. Hence, age gradients offer a test of current star formation and feedback models that can be probed via observations of resolved stellar populations.

\end{abstract}



\section{Introduction} \label{sec:intro}

Dwarf galaxies with $\Mstar \lesssim 10^8$ M$_{\odot}$ are observed to have younger stars closer to the center of the galaxy and older stars closer to the outskirts. This trend was first noticed by \cite{Tully_1996}, who saw that the least massive galaxies in the Ursa Major Cluster became redder with increasing radius, while the more massive galaxies in the cluster became bluer with increasing radius. 
We have since confirmed that this photometric color trend is found across a wide variety of dwarfs in both isolated and group environments \citep{Jansen_2000, Hunter_2006, Tortora2010, Liao2023}. Spectroscopic studies, resolved stellar populations, and multi-wavelength photometric studies have found that this color gradient corresponds to an age gradient in dwarf galaxies \citep[e.g.,][and references below]{Koleva2011, Zhang2012}, though a metallicity gradient can be present as well \citep[e.g.,][]{Harbeck2001, Tolstoy2004, Battaglia2006, Battaglia2008, Battaglia2011, Martinez2015, Martinez2021, Taibi2018, Han2020, Fu2024a, Fu2024b}.

The trend of the youngest stars at the center and the oldest stars dominating the outskirts has been seen in resolved stellar populations in nearby dwarf galaxies \citep[e.g.,][]{delPino2015, Okamoto2017, Bettinelli2019}. This trend is found across morphological types; in irregular \citep{Aparicio2000, Bellazzini2014, Gallart2008, Indu2011, Javadi2017, McQuinn2017, Piatti2018, Albers2019}, transition \citep{McConnachie2006, Hidalgo2009, Hidalgo2013}, and elliptical/spheroidal dwarfs \citep{Sharina2023, Bettinelli2019}. This is in contrast with Milky Way-mass galaxies, which tend to have gradients with blue, young stars on the outside and redder, older stars on the inside \citep{Tully_1996, Kepner_1999, Chiappini_2001}. 

We have a better understanding of how the age gradients of Milky Way-mass galaxies form compared to the gradients of dwarf galaxies, because the growth of disks in $\sim L^{\star}$ galaxies has been studied for many decades. Larger galaxies tend to form inside-out, with the first stars forming at the center of the galaxy out of collapsing low angular momentum gas and later stars forming in the outskirts of the disk as a result of gas accreted at later times with higher angular momentum \citep{Larson_1976, Williams2009, Gogarten2010, Barnes2014, Delgado2014, Delgado2015, Morelli2015, Dale2016, Zheng2017, Goddard2017, Peterken2020, Pessa2023}.  This inside-out formation is in agreement with disk formation theory, where the sizes of galactic disks increase with time as higher angular momentum gas is accreted 
\citep{WhiteRees1978, Fall1980}, and has also been seen in simulations of Milky Way-mass disks \citep[e.g.,][]{Sales2012, Zavala2016, Agertz2021, Bird2021, Ma2024}. As a result, inside-out formation explains the formation of stellar age gradients in Milky Way-mass galaxies to first order. 

In contrast, dwarfs tend to appear as though they have formed `outside-in,' with older stars less centrally concentrated and younger stellar populations more centrally concentrated. Theoretical explanations for this trend fall into roughly two groups: one explanation suggests that stars stay where they are formed (thus forming in an `outside-in' scenario). Another explanation could be that stars initially form `inside-out,' as in Milky-Way type galaxies, but are reshuffled over time. 

In order for the outside-in formation model to be a plausible explanation for the creation of dwarf galaxy age gradients, star formation in the outer regions of dwarfs would need to be suppressed, while stars continue forming in the inner regions.  Suppressing star formation would likely require either a depletion of gas or a decrease in gas density in the outskirts. 
If a satellite dwarf is able to retain gas until infall to a parent halo, simulations have shown that tides can create bars in dwarfs that funnel gas to the central regions, while ram pressure removes gas from the halo \citep{Mayer_2006}.  This scenario could leads to `outside-in' star formation as the gas content of the dwarf is altered \citep[e.g.,][]{Kawata2006, Stinson_2009, Genina2019}.  
Furthermore, the host halo environment itself can suppress gas accretion onto the satellite \citep[strangulation, e.g.,][]{Prescott2011}, which may cause the radius of star formation to shrink with time as gas is consumed.  

However, ram pressure and strangulation are primarily relevant to satellites, and do not explain the `outside-in' age gradients seen in field dwarf galaxies. \citet{Elmegreen2014} found that constrained accretion of gas (with the accretion rate matching the star formation rate) could explain the outside-in gradients of blue compact dwarfs, though it is unclear whether all dwarf age gradients could be explained in this way. Reionization and/or intergalactic UV radiation could potentially stop star formation in the outer regions of dwarf galaxies, where the gas is not dense enough for self-shielding, while still allowing for continued star formation in the center of the galaxy \citep{Gnedin2012}. Yet \cite{Zhang2012}, through studying multi-band surface brightness profiles of 34 nearby dwarf irregular galaxies with resolved HI profiles, found that dwarfs with an outside-in gradient have high enough HI column densities for self-shielding in their outskirts, implying that UV radiation would be unable to suppress star formation in the outer regions of field dwarfs. 

Given the challenges to outside-in formation, an alternative explanation is that the stars in dwarf galaxies get reshuffled with time. However, the mechanisms thought to cause stellar reshuffling in Milky Way-like galaxies may not be applicable in dwarfs. In large, spiral galaxies, bars and spiral arms are the main mechanisms that cause stars to migrate \citep[e.g.,][]{Hohl1971, Sellwood2002, Debattista2006, Roskar2008, Daniel2015, Frankel2018}.  While some disky dwarfs show evidence for stellar migration \citep[e.g.,][]{Radburn-Smith2012}, 
irregular or spheroidal type galaxies do not necessarily have bars or spiral arms to migrate their stars, and are unlikely to explain the frequency of the outside-in age pattern in field dwarfs. These considerations lead us to consider other causes of stellar reshuffling specific to dwarf galaxies. 

Several mechanisms have been shown to cause stellar reshuffling in simulated dwarfs, including mergers and fluctuations of the gravitational potential well that also lead to dark matter core formation, both of which are discussed below. \cite{BenitezLlambay2016} used cosmological simulations of Local Group-like galaxies and showed that dwarf-dwarf mergers could reproduce observed dwarf age gradients. In this model, they examined simulated dwarf Spheroidals (dSphs) and found that if a galaxy forms its older stellar population early, before reionization, then the energy from feedback and reionization can cause the gas to heat up and cease forming stars. Then, a dwarf-dwarf merger occurring after reionization could cool the gas and reignite star formation in the center of the halo while simultaneously causing the older stellar population to expand, resulting in an outside-in gradient. Other work done with simulations show that mergers can either flatten or steepen age gradients in isolated dwarf galaxies \citep{Graus_2019}. Specifically, if a galaxy has an initially flat age gradient but has a late time merger with a galaxy with an extended star formation history (SFH), incoming gas from the accreted galaxy forms new stars on the outskirts, thereby preventing an outside-in gradient and resulting in a flat age gradient. On the other hand, mergers can create a gradient if the galaxy merges with multiple small objects and old stars are added to the outskirts of the halo \citep[see also][]{Genina2019, Mostoghiu2018}.  Overall, it is not clear if mergers are common enough in the history of dwarf galaxies to explain the prevalence of outside-in age gradients \citep{Deason2014, Fitts2018}.

Another mechanism found in simulations that can reshuffle the stars in dwarf galaxies is stellar feedback. With stellar feedback, there are two mechanisms that previous authors have suggested can reshuffle the stars in a galaxy. The first is that bursty star formation can lead to bursty galactic winds, creating fluctuations in the shallow gravitational potential of the galaxy that can reshape the dark matter density profile from being steeply rising (a cusp) to flatter \citep[a core, i.e.,][]{Navarro1996, Read2005, Governato2010, Governato2012, Pontzen2012, Teyssier2013, DiCintio_2014, Brook2015, Chan_2015, Lazar_2020}, in line with the observed densities in dwarf galaxies \citep[e.g.,][]{Oh2011, Teyssier2013, Relatores2019}.  These fluctuations reshuffle both dark matter and stars.  The progression of the dark matter outward creates a dark matter core, while the stars are reshuffled such that the oldest stars move toward the outskirts of the galaxy over time \citep{Pontzen2012, ElBadry_2016, Graus_2019, Burger2022}. Thus, it is possible that age gradients are created via the same processes that create dark matter cores. 

The second mechanism by which stellar feedback can reshuffle stars in simulations is subtly different. Specifically, \cite{Stinson_2009}, \citet{Maxwell2012}, and \cite{ElBadry_2016} find that feedback causes outflowing (and later infalling) gas that can be star forming.  This cycle of inflow/outflow has sometimes been referred to as a ``breathing mode'' in dwarfs. Thus, newborn stars formed from this out/inflowing gas have an inherent initial radial velocity, causing them to migrate outwards. \cite{ElBadry_2016} finds that stars move $\sim$1 kpc in the first 100 Myr after they are born. 
\cite{Stinson_2009} also finds that the envelope within which star formation occurs contracts as the gas supply shrinks, reducing pressure-support and causing younger star formation to be limited to the center of the galaxy. 

All of the above simulation studies tend to focus on a specific type or mass of dwarf, e.g., field dwarfs only, or dwarf spheroidals only. 
This paints an incomplete picture of what causes the age gradients that are seen in dwarfs across morphological types and environment. So far, no study has systematically examined the effect of both environment and mass over a wide range of masses on the age gradients of dwarf galaxies. Using our sample of 73 simulated dwarfs with a mass range between $10^{5.8}-10^{9.6} \Msol$, we triple the sample used in the most recent previous work, \cite{Graus_2019}, and also look at satellite and central/field galaxies. We explore how dark matter core creation, star formation history, merger history, and environment combine to play a role in creating the age gradients of dwarf galaxies, and we attempt to disentangle the relative contributions of these effects. In Section 2, we describe the simulations used in our analysis and how we selected the galaxies in our sample. In Section 3, we present the methods used to analyze the simulated galaxies and the results of this analysis. Section 4 is where we discuss the implications of our findings. 
We summarize the key results in Section~\ref{sec:conclusion}.

\section{Simulations and Data}\label{sec:sims}

The simulations we use are run with \textsc{ChaNGa} \citep{Menon_2015}, an N-Body tree + Smoothed Particle Hydrodynamics (SPH) code based on its precursors \textsc{Gasoline} \citep{Wadsley2004} and \textsc{PKDGRAV} \citep{Stadel2001}. We use two suites of simulations, each using the "zoom-in" volume renormalization technique to resimulate pre-selected galaxies at a high resolution \citep[e.g.,][]{Katz_1993, Onorbe2014}. In order to implement photoionization and heating rates, a cosmic UV background is implemented \citep{Haardt2012}.
The star formation recipe used in the simulations is described in full in \cite{Christensen_2012}. This recipe uses temperature and density criteria in order to form stars (stars can form if the gas density is $\rho > 0.1 \text{cm}^{-3}$ and $T < 1000$K), but also weights the probability of forming stars by the fraction of molecular hydrogen, $H_2$, present. The requirement that $H_2$ be present leads to most star particles forming at densities $\rho > 100 \text{cm}^{-3}$. The probability $p$ of forming star particles in some time range $\Delta t$ is given by
\begin{equation}
    p = \frac{m_{\text{gas}}}{m_{\text{star}}}(1 - e^{-c_0^* X_{H_2}\Delta t/t_{\text{form}}}),
\end{equation}
where $m_\text{gas}$ and $m_{\text{star}}$ are the masses of the gas and star particle, respectively, $X_{H_2}$ is the mass fraction of molecular hydrogen in the gas particle,  $t_{\text{form}}$ is the local dynamical time, and $c_0^*$ is  the star formation efficiency parameter, set to $c_0^* = 0.1$. 
Each star particle in the simulation is represented with a Kroupa IMF \citep{Kroupa2001}, and metal line cooling and diffusion of metals is also included as described in \cite{Shen2010}.
The supernova (SN) feedback recipe is implemented based on the blastwave model from \cite{Stinson_2006}. This recipe releases $1.5 \times 10^{51}$ erg of thermal energy per SN. The energy deposited into the interstellar medium (ISM) represents the energy deposited by all processes related to young stars, including UV radiation from massive stars \citep{Wise2012, Agertz2013}, in addition to SNe.  In order to match the theoretical timescale of the snowplow phase of SNe\,II \citep{McKee1977}, cooling of gas particles is disabled for the lifetime of the momentum conserving phase. Cooling is not disabled for SNe\,Ia. SNe also deposit metals into nearby gas.

We look at eight total simulations, divided into two different simulation suites. The first suite are the Marvel-ous dwarfs, a group of four simulations (named Captain Marvel, Elektra, Rogue, and Storm) that yield a total sample of 68 field dwarfs from lower-density environments \citep{Munshi_2021, Christensen2024}. Although traditional zoom-in simulations select one halo of interest, and then place the highest resolution particles on that halo, the Marvel simulations select regions of the universe with dozens of dwarf galaxies, then re-run the entire region with high resolution.
Each of the Marvel simulations represents a (25 Mpc)$^3$ volume of the universe and uses a WMAP3 cosmology \citep{Spergel_2007}, though the high-resolution zoom regions are roughly $\sim$1 Mpc in diameter. The halos are typically 1.5-10 Mpc away from a Milky-Way mass galaxy, so these simulated dwarfs are representative of field dwarfs in the Local Volume \citep[see][]{Christensen2024}. The force softening resolution is 60 pc, and the dark matter particle masses are 6660 $\Msol$, the gas particle masses are 1410 $\Msol$, and star particle masses are 422 $\Msol$ at birth. 
Another suite used are the Near Mint DC Justice League (DCJL) simulations, a group of four simulations named after the first four female United States Supreme Court Justices \citep[Sandra, Ruth, Elena, and Sonia;][]{Akins2021, Bellovary_2019}. To create these simulations, environments near (out to $\sim$1 Mpc) Milky Way-mass disk galaxies were re-simulated at a higher resolution in order to capture satellite dwarf galaxies and nearby dwarfs. As implied, each of these simulations contains a central Milky Way-mass galaxy with a virial mass between $\sim 10^{11.9}-10^{12.4}$ M$_{\odot}$. Each of the DCJL simulations represents a (50 Mpc)$^3$ volume of the universe and uses Planck cosmological parameters \citep{Planck2016}. The force softening resolution is 170 pc, with particle masses of $4.2\times10^4~\Msol$ for dark matter, $2.7\times10^4~\Msol$ for gas, and $8000 ~\Msol$ for stars at birth. 

All halos in both suites of simulations are found using the Amiga Halo Finder  \citep[AHF;][]{Knollmann_2009}. Halos are identified using the overdensity criterion for a flat universe \citep{Gill_2004}, and the virial radius of a halo is defined as the radius for which the average halo density is  $200 \times \rho_{\text{crit}}(z)$, the critical density of the universe at a given redshift.
After running AHF, we identify a halo as resolved if it has at least 1500 dark matter particles and an extended star formation event lasting $\geq 100$ Myr. 

These and previous generations of simulations with these models have shown a number of successes in reproducing observations, including reproducing the observed stellar mass -- halo mass relation \citep{Munshi2013, Munshi_2021, Christensen2024}, the satellite galaxy distribution of massive galaxies \citep{Zolotov2012, Brooks2014}, the Kennicutt-Schmidt relation \citep{Christensen_2012}, the mass-metallicity relation \citep{Brooks2007, Christensen2016}, and properties of dwarf galaxies in both the Local Group and Local Volume \citep{Christensen2024}.

\subsection{Simulated Sample}\label{sec:sample}

Although there are 165 total resolved galaxies in the Marvel and DCJL simulations, we adopt a slightly stricter criteria when selecting galaxies for our sample since our analysis of the age gradient requires close dissection of the inner structure of each galaxy. We want to ensure that the age gradients are resolved and that the results of our analysis are not impacted by numerical effects. The density profiles at the center of simulated galaxies tend to converge at some length greater than a multiple of the gravitational softening length, referred to as the smallest convergent radius.  Simulations tend to suffer from artificial two-body relaxation on scales smaller than the smallest convergent radius \citep{Power_2003, Ludlow2020}. Thus, analyses of properties using lengths less than the smallest convergent radius may have numerical errors. \citet{Power_2003} finds that the smallest convergent radius is $\gtrsim 0.005 r_{200}$, while \citet{Ludlow2020} finds that the convergent radius occurs $\gtrsim 0.055 \frac{L}{N_p}$, where $L$ is the simulation box length and $N_p$ is the number of particles in one dimension. \cite{Zolotov2012} analyzes the convergence of earlier generations of the Marvel and DCJL simulations used in this work, with similar star formation and feedback implementation, and finds that the smallest convergent radius occurs within 2-4$\times$ the gravitational softening length. We adopt this criteria for our analysis, considering length scales below 2-4 times the force softening length to be unresolved.  $0.25$ kpc is 4$\times$ the softening length in the Marvel simulations and 2$\times$ the softening length in the DCJL simulations. Thus, we consider any halo with a 2D projected effective radius less than $0.25$ kpc to be unresolved, and remove these halos from our sample. After applying this criteria, we are left with $73$ total halos in our analysis, with Marvel's smallest galaxy having $\Mstar = 10^{5.26}~\Msol$ and 1166 star particles and DCJL's smallest galaxy having $\Mstar = 10^{5.81} ~\Msol$ and 219 star particles. 

Halo mass and stellar mass of all 73 simulated dwarf galaxies are listed in Table \ref{Table1} of Appendix \ref{AppA}.  All stellar masses reported in this work have been reduced from the full stellar mass determined by AHF based on the mock observational results in \citet{Munshi2013}.  In summary, a photometric analysis of mock photometry yields a stellar mass 0.6 lower than found by AHF. 

\subsection{Observational Data}\label{sec:obs_sample}
In Figs. \ref{fig:dt90_t90_overview} and \ref{fig:dt50_all} we also overplot observed radial age gradients (and uncertainties) for three observed galaxies: Leo A (Center), NGC-4449, and DDO-210.  These values represent a preliminary subset of results from a larger sample of $\sim$40 Local Volume gas-rich dwarf galaxies detailed in a forthcoming study (R.~E.~Cohen et al., in prep.).  Briefly, each galaxy was divided spatially into elliptical annuli, and a lifetime star formation history was fit to the color-magnitude diagram of stars in each annulus independently.  The radial age gradient slope and its uncertainties were then calculated using maximum-likelihood fits of a line to the individual per-annulus values of $t_{50}$ and $t_{90}$ (the times when 50\% or 90\% of the cumulative stellar mass was formed) versus radius, normalized to the observed half-light radius of each galaxy.  

\section{Results}\label{sec:results}
\subsection{Age Gradients}\label{sec:gradients_def}
To examine age gradients in this paper, we choose an analysis that can be carried out using resolved stellar populations, and compared against the observational data described in Section \ref{sec:obs_sample}.  Specifically, we measure the quantities $t_{50}$ and $t_{90}$, the times when 50\% or 90\% of the cumulative stellar mass was formed.  The values $t_{50}$ and $t_{90}$ can be determined from cumulative SFHs derived from deep color-magnitude diagrams.  We examine these quantities both as a function of radius in each galaxy, and globally for each individual galaxy.  A similar measurement was adopted by \cite{Graus_2019} for the {\sc fire}-2 galaxies, which enables a direct comparison to their results (see here and in Section \ref{sec:t50_results}). 


First, we compute the 2D projected effective radius with the galaxy oriented face-on, meaning that the total angular momentum vector (the sum of the angular momentum of all the particles identified by AHF to be bound to the galaxy) is aligned with the $z$-axis. Next, we calculate the total $V$-band luminosity of the galaxy using {\sc pynbody}, which adopts the Padova simple stellar population models \citep{Marigo2008, Girardi2010}, and determine which circular radius encloses half of that luminosity. This radius enclosing half the light is $R_e$, the circular 2D effective radius.

\begin{figure*}
    \centering \includegraphics[width=0.85\textwidth]{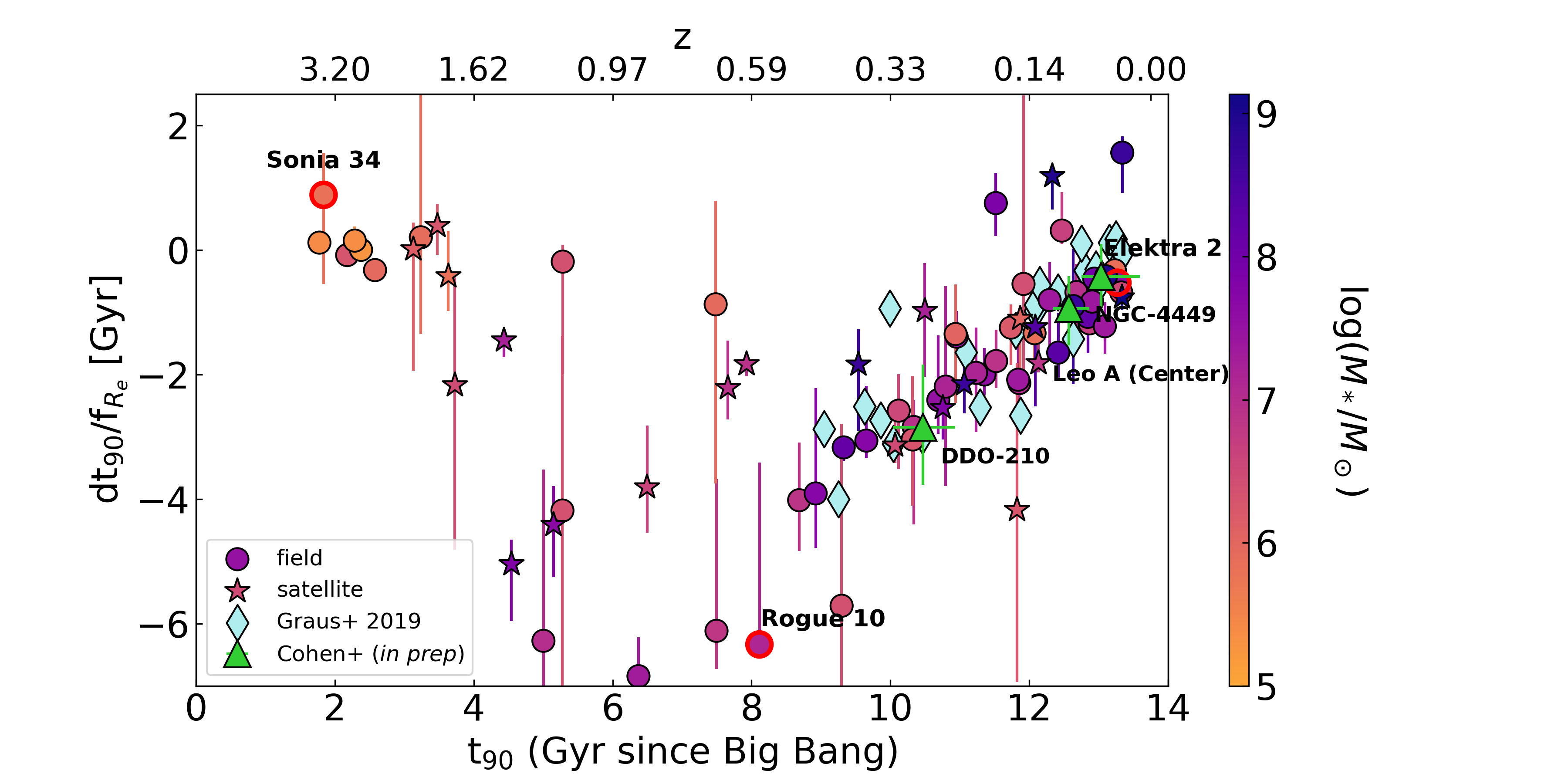}
    \caption{The normalized age gradient value $dt_{90}/f_{R_e}$ vs.~global $t_{90}$ in Gyr since Big Bang. Circles represent simulated field dwarfs and stars represent simulated satellite galaxies. Error bars on $dt_{90}/f_{R_e}$ enclose 68\% of the age gradient values calculated over 100 random viewing angles. The color of the points corresponds to the stellar mass of the galaxy in solar masses. Light blue diamonds correspond to the simulated field galaxies presented in \protect\cite{Graus_2019}, which have a stellar mass range of $10^{5.67-8.73}$ M$_{\odot}$. Green triangles correspond to three observed galaxies, a subset of galaxies currently being analyzed  (R.~E.~Cohen et al., in prep). We also label three galaxies we refer to as ``representative'' galaxies, which are presented in Fig.~\ref{fig:r_age_plots} and referenced throughout this work. 
    Similar to \protect\cite{Graus_2019}, we find that age gradients change from an outside-in trend (strongly negative $dt_{90}/f_{R_e}$) to no age gradient as the global $t_{90}$ increases after $\sim$8 Gyr, i.e., galaxies with more extended star formation histories have flatter age gradients. However, with our larger sample we show that this trend reverses as $t_{90}$ decreases. We also find that both field dwarfs and satellites follow similar age gradient trends. Both sets of simulations are in good agreement with the observational results.}
    \label{fig:dt90_t90_overview}
\end{figure*}

The next step in calculating the age gradient is taking all the stars within 1.5 times the effective radius and dividing the stars into ten annular bins such that each bin contains an equal number of stars. We calculate $t_{90}$, the time (since the Big Bang) when 90\% of the cumulative mass of stars ever formed have been formed, for the stars in each annular bin. We take the difference of the innermost $t_{90}$ value and outermost $t_{90}$ value in order to determine how the age of the stars in the central region of the galaxy differs from the age of stars at 1.5 $R_e$, and divide by the distance between the centers of the inner and outer bins. This difference is the $dt_{90}$ value. The $dt_{90}$ value is then normalized with respect to the effective radius $R_e$: 
\begin{equation}
    \centering
   \left. \frac{dt_{90}}{f_{R_{e}}} \right|_\theta = \frac{t_{90, \text{out}} - t_{90, \text{in}}}{(R_{\text{out}} - R_{\text{in}})/R_e}
    \label{eq:dt90}
\end{equation}
Where $f_{R_{e}} = \frac{R_{\text{out}} - R_{\text{in}}}{R_e}$ and $R_{\text{out}}$ and $R_{\text{in}}$ represent the center of the outer and inner radial bin, respectively. The $\theta$ denotes that this is the age gradient computed at a particular viewing angle, discussed in the next paragraph. Thus, we refer to the quantity $\left. \frac{dt_{90}}{f_{R_{e}}}\right|_\theta$ as the age gradient at a particular viewing angle.

We recalculate Equation \ref{eq:dt90} for 100 random viewing angles ($\theta_i$) for each galaxy in order to quantify bias introduced by observational viewing angle, and we take the median over all viewing angles to be the final age gradient. For each different sightline, the circular $R_e$ is recalculated (i.e., it is not always the face-on $R_e$ described above that is used):
\begin{equation}
    \centering
    \frac{dt_{90}}{f_{R_e}} = \frac{1}{100}\sum_{i=1}^{100}\left. \frac{dt_{90}}{f_{R_{e}}}\right|_{\theta_i}
    \label{eq:dt90_global}
\end{equation}

We refer to this quantity $\frac{dt_{90}}{f_{R_e}}$ as the stellar age gradient of the dwarf galaxy. Note that $\frac{dt_{90}}{f_{R_e}}$ has units of Gyr because it has been normalized to $R_e$.

In addition to the $t_{90}$ and $dt_{90}/f_{R_e}$ values, we also compute the $t_{50}$ and $dt_{50}/f_{R_e}$ values using a similar method by instead calculating the $t_{50}$ value, the time when 50\% of the stars were formed, within each bin. Similarly, $dt_{50}/f_{R_e}$ is the difference between the inner and outer $t_{50}$ value, normalized to the effective radius. The $t_{50}$ results are shown and discussed in Section \ref{sec:t50_results}. The global $t_{50}$ and $t_{90}$ values of each simulated galaxy are listed in Table \ref{Table1} of Appendix \ref{AppA}, along with their respective age gradients.   Table \ref{Table1} of Appendix \ref{AppA} also lists the face-on 2D $R_e$ and $V$-band magnitude of each galaxy, as well as whether the galaxy is a satellite or field dwarf. 

Age gradients as a function of global $t_{90}$ are presented in Figure \ref{fig:dt90_t90_overview}. We calculate global $t_{90}$ with the same method used to calculate $t_{90}$ in each radial bin described above, but instead use all of the star particles within the simulated galaxy. We denote isolated field galaxies as circles and satellite galaxies as stars, with satellite galaxies defined by AHF as halos within the virial radius of another, more massive halo. The color bar shows log stellar mass in units of solar mass, and we plot the global $t_{90}$ value of the galaxy in Gyr since the Big Bang on the $x$-axis and the stellar age gradient, $dt_{90}/f_{R_e}$ (see eq. \ref{eq:dt90_global}), on the $y$-axis. Error bars on $dt_{90}/f_{R_e}$ enclose 68\% of the age gradient values calculated over the 100 random viewing angles. Galaxies with a $t_{90}$ close to 14 Gyr are gas-rich and star forming at $z=0$, but the galaxies with a $t_{90} \lesssim 2 \text{ Gyr}$ were quenched soon after reionization. A $dt_{90}/f_{R_e}$ close to 0 means that the age gradient is flat, while a negative $dt_{90}/f_{R_e}$ value implies a steep, negative, `outside-in' looking age gradient. 
Two outlier galaxies with extremely negative age gradients 
are not shown on the plot\footnote{Two outliers are not shown on this plot because they lie outside of the $y$-axis range, being more than $2\times$ the standard deviation away from the mean age gradient of the ensemble of simulated galaxies. Both are galaxies with extremely negative age gradients less than $-6.56 \text{ Gyr}$ (Storm 14 and Ruth 3). We have examined these and discovered they form their stars at roughly the same radius and that the oldest stars migrate to the outskirts of the galaxy over time. For both galaxies, old stars ($\text{age } > 9\text{ Gyr}$) dominate the galaxy at all radii and make up at least 90\% of the stars beyond $\sim 2\times R_e$. Both of these galaxies have a small fraction (10-15\%) of stars with ages $< 1 \text{ Gyr}$ in the inner 0.25kpc radii of the galaxy and none beyond at least $2\times R_e$.}.  
In general, we see that galaxies that quenched early and galaxies that are still star-forming both have flat age-gradients, while the steepest (negative) age gradients occur for galaxies with a $t_{90}$ values near $\sim 6-8 \text{ Gyr}$. Both field and satellite dwarf galaxies follow this trend with no significant difference between them.

Also shown in Fig.~\ref{fig:dt90_t90_overview} are the simulated galaxies from \cite{Graus_2019}, which span roughly a similar range in stellar mass (4.5$\times10^5 - 8.5\times10^8$ M$_{\odot}$), but include only isolated, field dwarfs. 
We note two slight differences between our study and that in \citet{Graus_2019}.  First, \citet{Graus_2019} defines $t_{50}$ and $t_{90}$ as lookback times, while we use the time since the Big Bang. We have put the {\sc fire}-2 results on our plot by assuming an age of the Universe of 13.8 Gyr.  Second, \citet{Graus_2019} defines $t_{50}$ and $t_{90}$ in terms of the lookback time to the final stellar mass, while we quantify them in terms of the cumulative stellar mass ever formed.  That is, the \citet{Graus_2019} definition is in terms of the $z=0$ stellar mass, while we determine the cumulative SFH, which includes stellar mass from massive stars that are no longer around at $z=0$.  We have chosen the latter because it is the method commonly employed by observers.  We assume the difference in definitions are minor and compare our results to those in \citet{Graus_2019}.

We see that, in general, our simulations match the trend presented in \cite{Graus_2019} in the region where we have overlapping global $t_{90}$, and that both sets of simulations are in good agreement with the preliminary observational results. However, despite having a similar stellar mass range, our sample of 73 galaxies is much larger than \cite{Graus_2019}'s sample of 26 field galaxies and also includes satellite galaxies.  With our larger, more diverse sample of galaxies, we predict a "U"-shaped trend: galaxies that quench soon after reionization tend to have flat age gradients, and the gradients steepen as global $t_{90}$ increases between $t_{90} \sim 4-8 \text{ Gyr}$, until it reverses direction at $t_{90} \sim 8$ and gradients begin to flatten again with more recent $t_{90}$. 
The galaxies that quench soon after reionization tend to be the lowest masses (M$_\ast \lesssim 10^{6} ~\text{M}_\odot$) and the galaxies with $t_{90}$ close to $z=0$ tend to be the most massive (M$_\ast \sim 10^{9} ~\text{M}_\odot$).  Galaxies over this mass range 
exhibit a wide range of age gradients, which we explore further below.

\subsection{Inside-Out vs Outside-In Growth}\label{sec:reshuffling_def}

\begin{figure*}
    \centering
    \includegraphics[width=0.95\textwidth]{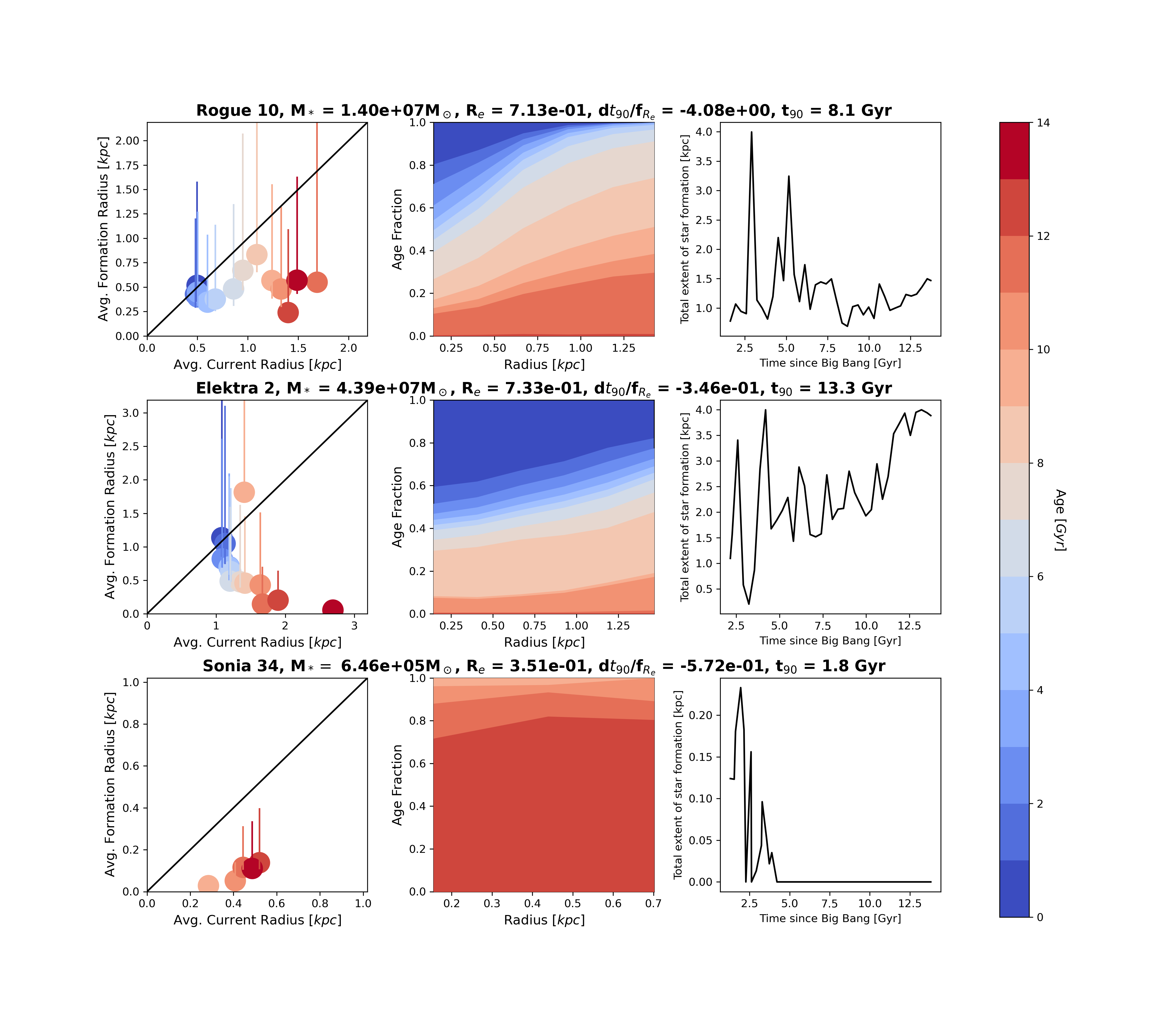}
    \caption{Examination of gradients in three representative dwarf galaxies. From top to bottom, they are Rogue 10, Elektra 2, and Sonia 34. For the left and middle columns, we bin the star particles by their age in 1 Gyr bins, with the  reddest color denoting star particles that formed 13-14 Gyr ago, and the bluest color representing stars that formed 0-1 Gyr ago. The left column shows average formation radius vs.~average radius at $z=0$ for the binned star particles in order to determine how the stellar positions have changed over time, with the vertical lines enclosing 90\% of the formation radii of each bin. The middle column shows fraction of stars of each age as a function of radial position at $z=0$. The rightmost column shows the extent of star formation from the center of the galaxy over time. 
    For all galaxies, the left plots show that the oldest stars have been pushed out the furthest from their initial formation radius. 
    Steep, negative age gradients result from little growth in the size of the galaxy over time, while the oldest stars get pushed beyond the mean formation radius.  Flat gradients result when the galaxy grows with time, such that young stars are forming at the radius that the oldest stars have been pushed to. The low mass galaxies have stars of similar (old) ages, and thus tend to have flat gradients due to a lack of evolution in star formation. 
    A full repository of the age gradient and radial reshuffling plots for each galaxy in our sample is available online at \dataset[10.5281/zenodo.13887797]{https://doi.org/10.5281/zenodo.13887797}.}
    \label{fig:r_age_plots}
\end{figure*}

A number of our simulated dwarf galaxies exhibit steep, negative age gradients, with younger stars in the center and older stars near the outskirts. However, determining if these stars form inside-out initially and have their stars reshuffled over time or if they form outside-in requires analyzing the stars' motion over time. In Fig.~\ref{fig:r_age_plots}, we examine three representative galaxies from our sample (which are also labeled in Fig.~\ref{fig:dt90_t90_overview}), each exhibiting a different type of age gradient, in order to demonstrate how their stars move over time. In the left-most column of Fig.~\ref{fig:r_age_plots}, we explore how the star particles' positions vary with time.  In all of these plots, redder coloring denotes older star particles, while bluer denote younger star particles.  We bin the stars in each galaxy by age (bin width = 1 Gyr) and plot their mean formation radius versus their mean $z=0$ radius. The vertical lines show the region enclosing 90\% of the star formation per bin. The 1-1 line shown in the left column represents where the stars would be if they hadn't moved from their radial position relative to the center of the galaxy at all, while points above this line imply that the stars in that bin moved radially inward on average, and points below this line mean that the stars in that bin tended to move radially outward relative to where they formed. 

In the middle column of Fig.~\ref{fig:r_age_plots} we show the age gradient at z=0 for each galaxy by plotting the fraction of stars of each age in each radial bin out to $2\times R_e$ for each galaxy. The rightmost column shows the total extent of star formation over the lifetime of each galaxy. The $x$-axis denotes the time since the Big Bang in Gyr, while the $y$-axis displays the furthest radius away from the center of the galaxy that contains a newly formed star particle at that time. 
 
As can be seen from the middle column, each of these age gradients is quite different. In fact, the three field galaxies shown in Fig.~\ref{fig:r_age_plots} were chosen because they each exemplify one of three general types of gradients we see in our sample. 
The top galaxy in Fig.~\ref{fig:r_age_plots} is halo 10 from Rogue, which exemplifies a galaxy with a steep, negative age gradient. The second is halo 2 from the Storm simulation, which is a more massive dwarf galaxy that has a relatively flat age gradient. 
Sonia 34 is a low mass dwarf with a flat age gradient composed of old stars, indicating it quenched its star formation early on.

For the galaxy with the steep stellar age gradient, Rogue 10, we see that all of the stars regardless of age formed at roughly 0.5 kpc from the center of the galaxy. However, at $z=0$, the oldest stars have migrated away from the center of the galaxy to $\sim$1.5 kpc, while the youngest stars remain closer to their initial position of 0.5 kpc. The migration of the oldest stars results in a steep, outside-in age gradient.  Elektra 2's flat gradient shows that the oldest stars formed closest to the center, while younger stars formed further and further away from the center; i.e., a clear inside-out formation trend (with the exception of the stars between 9-10 Gyr old). We also see that although the mean formation radius of the youngest star particles is around $\sim$1 kpc from the center, young stars form out to $\sim$3 kpc. Since young stars form across the halo, this ends up creating a flat gradient. For Sonia 34, a low mass dwarf with the flat age gradient, there are only old stars and we can see that it doesn't form any stars after 4 Gyr.  While the oldest stars do seem to end up further from their formation radius, indicating some reshuffling, the flat gradient is due to early quenching and a lack of continued star formation after 4 Gyr.

The pattern seen in Rogue 10 persists across the majority of our sample with steep age gradients: both old and young stars tend to form at roughly the same radius from the center, but the older stars have significantly migrated from their mean formation radius. There are a few exceptions, where galaxies with a steep age gradient have radial plots that look more similar to Elektra 2's, with progressively younger stars forming further out with time relative to the older stars. However, in these cases the gradient is weighted by the fraction of stars that form at a given time, such that there is a much higher percentage of young stars at the center of the galaxy relative to the outskirts, causing the negative age gradient. Additionally, in most of these cases the mean formation radius of the young stars at $z=0$ is still interior to the mean radius that the oldest stars have migrated to, yielding a strong age gradient.  

For the more massive dwarf with a flat age gradient (Elektra 2), we again see that the oldest stars form closest to the center and move to the outer region of the galaxy over time. However, the primary difference between the flat gradient and the steep gradient is where the young stars form. When looking at the formation radii of the stars in the galaxy with the flat gradient, a clear inside-out trend is seen: the average formation radius of the younger stars is much further out than the older stars.  Additionally, the mean $z=0$ formation radius is equal to or further out than the mean radius that the old stars have migrated to. This trend is ubiquitous across the galaxies with a flat gradient. 

Overall, these plots show that regardless of current age gradient and mass, the oldest stars in our dwarf galaxies tend to form in the central region of the galaxy and over time migrate to the outskirts of the galaxy (we investigate this migration below). Then, the determining factor in a dwarf galaxy's age gradient is where the younger stars form. In galaxies with steep age gradients, whatever process migrates the older stars has pushed them to beyond the mean formation radius of the stars at any time, yielding a steep outside-in gradient.  In the galaxy with the flat age gradient, the star formation radius instead increases to match the radius that the older stars are pushed to, yielding an overall flat age gradient. 
For low mass dwarfs that quench early, their gradients are flat because most of their stars are similar in age (old), even if some migration has occurred. 

\subsection{Reshuffling}\label{sec:reshuffling}
There are two mechanisms that can reshuffle stars, discussed in-depth in Sec.~\ref{sec:intro}. While stellar feedback from bursty star formation drives both mechanisms, they differ subtly from each other. One way that stellar feedback can reshuffle stars is by causing radial flows in star forming gas, such that when stars form they have an inherent radial velocity which causes some to migrate outwards \citep{Stinson_2009, ElBadry_2016}. The other mechanism through which stellar feedback can cause stellar reshuffling is by fluctuating the shallow gravitational potential of the galaxy, a process that has been shown to lead to flattening of the dark matter density profile \citep[e.g.,][]{Pontzen2012, ElBadry_2016}. Both the stars and dark matter are collisionless particles, and the fluctuations can cause both stars and dark matter to move outward over time. However, it is important to note that these two mechanisms are not necessarily mutually exclusive.  In fact, they both occur in the simulations of \citet{ElBadry_2016}. We examine both the radial velocity of stars over time and the dark matter core slope in order to assess the impact of each of these process in reshuffling stars in our simulations. 
\subsubsection{Stellar Radial Velocities at Birth}\label{sec:breathing}
\begin{figure*}
    \centering
    \includegraphics[width=0.95 \textwidth]{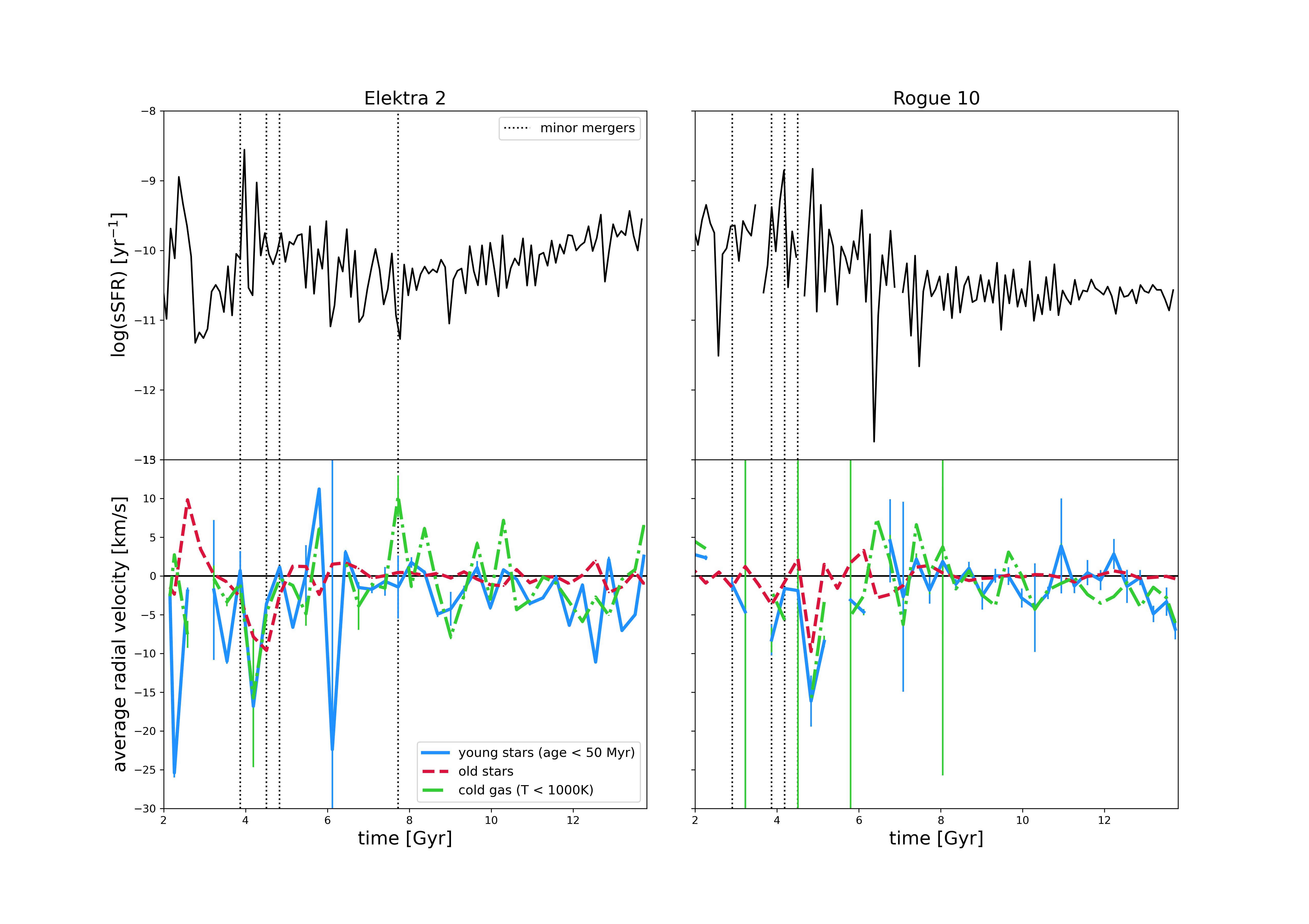}
    \caption{The top two panels show the specific star formation rate (sSFR) for Elektra 2 (left) and Rogue 10 (right) and the bottom two panels show the average radial velocity over time of old stars (dashed red line), young stars (solid blue line), and cold gas (dot-dashed green line) for these galaxies. Old stars are defined as having an age greater than 2 Gyr at the timestep at which the radial velocity was calculated, while young stars have an age less than 50 Myr at the timestep at which the radial velocity was calculated. Cold gas is gas with temperature $T < 1000$ K. Vertical dashed lines indicate when a minor merger occurred in each of these galaxies (neither galaxy experienced a major merger). Error bars are shown for all lines, which represent 
    $\sqrt{N}/N_{\rm bin}$ so that bins with less particles have a greater error. 
    While we tend to see larger fluctuations in the radial velocities of stars and gas during the period of minor mergers, the outward radial velocities are substantially smaller in our simulations than those found in \citet{ElBadry_2016}, who found outward velocities of 20 km/s.  We find mean radial velocities near zero outside of minor mergers, or even a small net inflow.  The oldest stars show no evidence of a large outward radial motion that could drive the creation of the age gradients we see in these same galaxies.}
    \label{fig:radial_migration}
\end{figure*}
We investigate if our star particles inherit an initial radial velocity from their parent gas particles at birth.  \citet{Stinson_2009}, \citet{Maxwell2012}, and \citet{ElBadry_2016} found that their bursty star formation induced oscillating inflows and outflows in their star forming gas, a process often dubbed ``breathing.''  This breathing mode can induce at least some of the stars to have large outward radial velocities at birth.  To investigate whether a breathing mode can explain the stellar reshuffling we see in our simulated galaxies, in Fig.~\ref{fig:radial_migration} we look at two of our galaxies that were also shown in Fig.~\ref{fig:r_age_plots}, Elektra 2 and Rogue 10, our examples of a flat and steep gradient, respectively.  We use our simulated data to create plots similar to figure 3 from \cite{ElBadry_2016} but extend the time frame over the entire life of the galaxy rather than looking only at a 1 Gyr time interval. 
The top two panels plot the specific star formation rate (sSFR) in years$^{-1}$, i.e., the star formation rate in 100 Myr bins, normalized by the stellar mass of the galaxy at that time. Vertical dotted lines depict where minor mergers occurred in the halos' histories, as neither galaxy had major mergers. We define a minor merger as a merger event having a virial mass ratio less than 1/4. The bottom panels show the average radial velocity with respect to the center of the halo for old stars, young stars, and cold gas, where old stars are defined as star particles with ages greater than 2 Gyr at each timestep, young stars have an age less than 50 Myr at each timestep, and cold gas has a temperature less than 1000 K. Because star formation information for the particles is stored more often than the snapshots (which store the position and velocity information of the particles), we are able to plot sSFR with finer time resolution (100 Myr) compared to the radial velocities ($\sim$300 Myr). 

Error bars are shown for the all lines as vertical lines, where the error reflects $\sqrt{N}/N_{\rm bin}$ so that bins with less particles have a greater error. Some errors are quite large for the cold gas and young star samples, since the number of particles per bin in Elektra 2 ranges from 0-3324 for young stars and from 0-13735 for cold gas particles. For Rogue 10, similar values are found, with the number of star particles per bin ranging from 0-4216 and the number of cold gas particles per bin ranging from 0-2187. However, the errorbars are barely visible for the old stars, since for both simulations the number of old stars per bin is consistently large; for Elektra 2, the number of old stars per bin ranges from 3635-4669 and for Rogue 10 this number ranges from 16863-23760. 


In general, the stars in the simulations should inherit the velocities of their parent gas particles.  We have verified (not shown here) that the velocity dispersion of the young stars matches that of the cold ($T < 1000$ K), star-forming gas.  The more extreme inward flows during the time of minor mergers make this more obvious: the inward radial velocities of the young stars and cold gas are identical at those times.  During other times, this may not be as obvious because the cold gas is an instantaneous snapshot of radial velocity, while the young stars trace the mean cold gas radial velocity during the preceeding 50 Myr.  

Both galaxies show inflowing gas and stars with -15 km/s to -20 km/s at times, particularly during periods when minor mergers are occurring.  These times are often also associated with peaks in the sSFR.  In contrast, the outward radial velocities are never higher than 10 km/s for any of the three components.  Other than one burst in Elektra 2, the young stars are generally consistent with no net radial velocity, or even a slight net inward migration.   
In contrast, in the 1 Gyr time period \cite{ElBadry_2016} looks at, they find radial velocities for young stars, old stars, and star forming gas all oscillating from roughly -20 to 20 km/s, with peak-to-trough variations taking roughly 200-300 Myr. 
They also find that the peaks in radial velocity seemingly respond to the peaks in sSFR. 
Outside the time of minor mergers, we do not see the large variations in radial velocity seen in {\sc fire}.  However, our output time resolution is not as fine as in {\sc fire}, so it is possible that we might miss trends occurring on timescales smaller than our output timesteps, which is $\sim$300 Myr.  Despite that, our technique samples young stars (those that formed within the past 50 Myr) every $\sim$300 Myr.  We do not expect any bias against finding radially outflowing young stars in this sampling.  Likewise, it is clear that stars are not displaying any net outward motion as they age. 
Thus, we conclude that there is not significant evidence that stars are forming in outward radial gas flows.  This is in contrast to other dwarf galaxy simulations.  To fully understand whether there is any breathing mode occurring in our simulations will require further investigation, which we leave to future work. However, there is no immediate evidence that stars have a net radial velocity that could create the age gradients we find.

\subsubsection{Potential Well Fluctuation}
\begin{figure*}
    \centering
    \includegraphics[width=\textwidth]{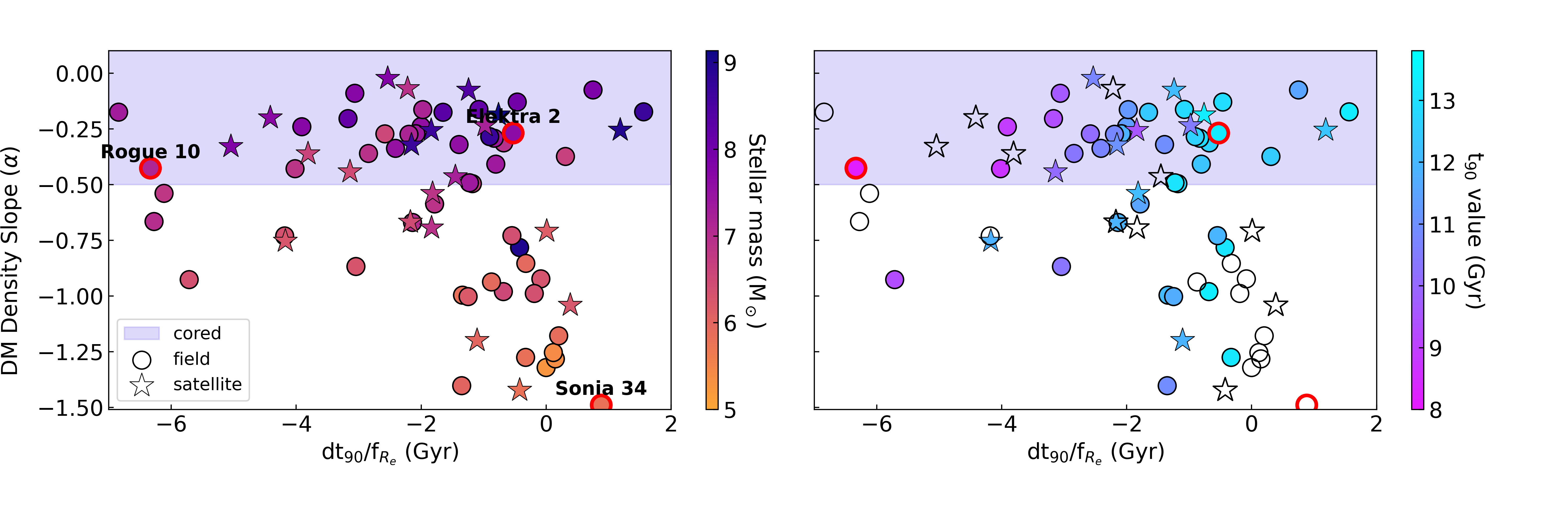}
    \caption{Both panels show dark matter density slope vs.~age gradient ($dt_{90}/f_{R_e}$) for simulated galaxies. Stars denote satellite galaxies while circles denote field galaxies.  The left panel is colored by stellar mass. In the left panel, it is evident that core slope varies as expected with stellar mass: halos with lower stellar mass are `cuspy,' while increasing mass leads to a `cored' profile. This trend extends somewhat to age gradient; lower mass `cuspy' halos have flat age gradients, while higher mass 'cored' halos have some of the steepest negative age gradients. The right panel focuses on the mid- to high-mass halos, and colors by global $t_{90}$ value. Empty symbols with no color are galaxies with $t_{90} < 8\text{ Gyr}$. Here it shows that amongst `cored' halos, those with the steepest age gradients tend to have a $t_{90}$ value closer to $\approx 8\text{ Gyr}$, while the flatter age gradients tend to have $t_{90}$ closer to $z=0$. The example galaxies shown in Fig.~\ref{fig:r_age_plots} are outlined in red in both panels and labeled with text in the left panel.  Overall, we interpret these trends as evidence that dark matter core formation correlates with age gradient for the lower-mass galaxies, but that for the higher-mass dwarfs with cores, $t_{90}$ instead dominates the age gradient.}
    \label{fig:core_dt90}
\end{figure*}
Because dwarf galaxies have a shallower gravitational potential well relative to larger, Milky Way-mass galaxies, stellar feedback can fluctuate the gravitational potential and thereby change the distribution of dark matter in a galaxy from having a sharply increasing inner slope (a cuspy profile) to a flatter inner slope \citep[a core, e.g.,][]{Pontzen2012}. Thus, to explore any correlation between core formation and reshuffling of stars, we calculate dark matter core slopes using the core-Einasto profile presented in \citet{Lazar_2020}. The inner-slope is calculated by fitting a line to the dark matter density profile between $1-2 \%$ of the virial radius $R_{\text{vir}}$.\footnote{If 0.01$R_{\text{vir}}$ was $<$ 250 pc (the lowest resolved radius, 2-4 $\times$ the gravitational softening length), then 0.25 kpc was used as the inner radius.  Otherwise, 1\%. } The slope of this line, $\alpha$, is the dark matter density slope referred to in Fig.~\ref{fig:core_dt90}, where we plot the dark matter density slope vs.~age gradient (quantified as before as $dt_{90}/f_{R_e}$, see \ref{sec:gradients_def}). Traditionally, cuspy density profiles (such as those found in dark matter-only simulations) tend to have $\alpha \leq -1$. Baryonic feedback has been shown to flatten $\alpha$. As seen in Fig.~\ref{fig:core_dt90}, there is a significant population of simulated galaxies with $\alpha > -0.5$ (highlighted by the shaded region), demonstrating that these halos have undergone dark matter core creation due to baryonic feedback. When baryonic physics creates dark matter cores in dwarf galaxies, the galaxies initially form cuspy halos at high $z$ before being transformed into cores at lower $z$ through extended star formation and feedback \citep[e.g.,][]{Fry2015,Muni2024}.  

We investigate whether or not feedback is changing the distribution of dark matter and thereby causing the stellar gradients we see in our simulated dwarfs by looking for a correlation in the stellar age gradient ($dt_{90}/f_{R_e}$) with respect to the dark matter density slope ($\alpha$) in Fig.~\ref{fig:core_dt90}. The left and right panel differ in how the points are colored: the left panel colors each point by its $\log_{10}(M_*)$ while the right panel colors galaxies with a global $t_{90}$ value greater than 8 Gyr since the Big Bang (and leaves halos with a $t_{90} < 8 \text{ Gyr}$ uncolored). The three halos from Fig.~\ref{fig:r_age_plots} 
are outlined in red and labeled. Looking at the left panel, we can see that the dark matter density slope varies with stellar mass; that is, galaxies with low stellar mass have cuspy dark matter density profiles while the highest mass have a cored dark matter density profile \citep[see also][]{Governato2012, DiCintio_2014, Tollet2016, Lazar_2020, Jackson2024, Azartash2024}. We also see that this trend generally holds true for both field (circles) and satellite (stars) galaxies. 

Fig.~\ref{fig:core_dt90} shows evidence for two trends: a diagonal one from top left to lower right and a horizontal one across the top of the graph.  The diagonal trend follows the mass trend.  That is, looking at the left panel we can see that the lowest mass galaxies  have cuspy dark matter halos and only have flat age gradients.  However, as we look at increasingly more massive galaxies (from $\Mstar \sim10^6$ to $10^8$ M$_{\odot}$), the dark matter density slopes become more shallow, going from roughly $-1.25$ to $-0.25$.  As this happens, the galaxies with dark matter cores are able to form steeper age gradients compared to their cuspier counterparts, and although they still form flat gradients (see the subsequent paragraph for more discussion on this point), they also form the steepest age gradients of $\sim$-6 Gyr. In summary, the cuspiest, lowest mass halos have flat age gradients while increasingly massive, cored halos have increasingly steep gradients. 
We interpret this trend as a signature of core creation, which becomes more significant with increasing stellar mass (or, vice versa, less significant with decreasing stellar mass).  The low-mass galaxies with cuspy profiles tend to have only flat stellar age gradients, but as the dark matter density slope flattens and becomes more cored with increasing stellar mass, the galaxies are able to develop steeper and steeper age gradients. We conclude that this is evidence that the same process that creates dark matter cores is also influencing the age gradients in our simulated galaxies.  In Appendix \ref{AppB} we verify the role of core formation in setting age gradients by examining a simulated galaxy in the case that it does or does not create a dark matter core, and find only the galaxy with core creation reshuffles stars and develops a steep, outside-in age gradient. 

However, as stellar mass continues to increase, the galaxies maintain cores and there is no longer a relation between core slope and age gradient.  Among the most massive, cored halos, the age gradients range from steep to flat, suggesting that a fluctuating gravitational potential due to stellar feedback is not the only mechanism at play in determining a dwarf galaxy's age gradient. Instead, we see evidence for a second trend that shows up in the cored galaxies (those with dark matter density slopes $< -0.5$).  This can be seen as a color gradient in the cored galaxies in the right panel of Fig.~\ref{fig:core_dt90}, which have  $t_{90} > 8 \text{ Gyr}$.  Steeper age gradients (near $-4$ Gyr) correlate with a $t_{90}$ closer to 8 Gyr (purple/magenta color), whereas flatter age gradients correlate with more recent star formation (cyan points). If we recall the discussion of Fig.~\ref{fig:r_age_plots}, flat gradients result from inside-out growth. Thus, for the massive dwarf galaxies with cored dark matter density profiles, the longer that star formation continues, the further out the galaxy forms stars and flattens gradients.  On the other hand, those more massive dwarfs with earlier $t_{90}$ seem to develop a steep gradient due to core formation, and maintain that gradient due to a lack of subsequent growth in size. We explore this further in the next section.

\begin{figure*}
    \centering
    \includegraphics[width=0.90\textwidth]{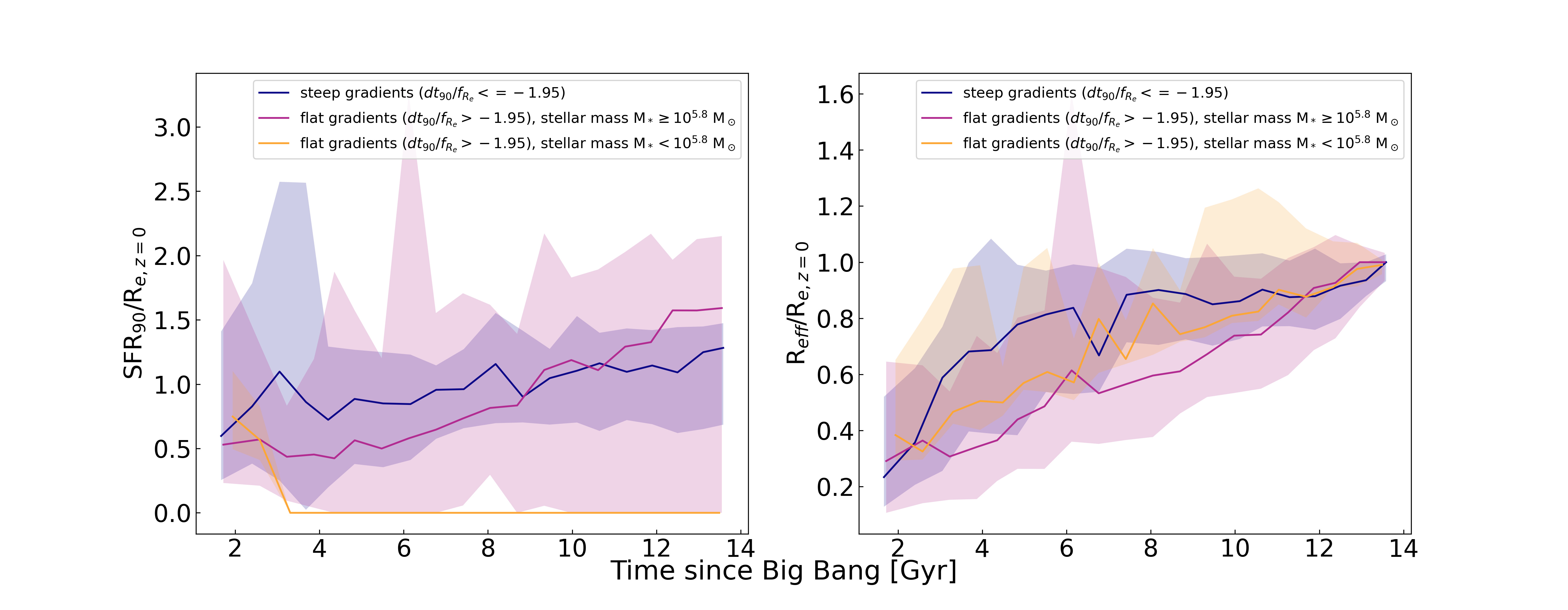}
    \caption{The left panel shows the average radius enclosing 90\% of the star formation over time normalized to the effective radius at $z=0$ while the right panel shows the effective radius normalized to the effective radius at $z=0$. We exclude satellites from this plot because their size evolution may be impacted by their evolution within a parent halo. Field galaxies are split up according to their age gradient value. 
    We define a flat gradient as a gradient with a $dt_{90}/f_{R_e}$ value above the average $dt_{90}/f_{R_e}$ value of our sample, $-1.95 \text{ Gyr}$, while a steep gradient is a gradient with a $dt_{90}/f_{R_e}$ value less than or equal to $-1.95$ Gyr. Furthermore, the sample of field galaxies with flat age gradients is split into high mass and low mass, with a cutoff at $10^{5.8}$ M$_\odot$. This is because smaller galaxies with a flat gradient have a flat gradient because they quench early, not because recent star formation erases their gradient. 
    In both panels, the shaded lines show the data within 68\% of the average value. 
    Overall, both panels show evidence that massive dwarfs with flat age gradients at $z=0$ started off smaller than their counterparts with steep gradients but overtook them more at recent times.
    }
    \label{fig:sf_history}
\end{figure*}

\subsection{Star Formation History and Size}\label{sec:sfh}
Fig.~\ref{fig:core_dt90} shows that galaxies with dark matter cores can span a range of age gradients, and the right panel shows that for these same galaxies, the age gradient varies with $t_{90}$ value. This motivates us to explore the star formation history of these dwarfs. The left panel of Fig.~\ref{fig:sf_history} shows the average radius enclosing 90\% of the star formation (denoted $\text{SFR}_{90}$) over time, normalized to the effective radius at $z=0$. The right panel shows the history of $R_e$ normalized to the effective radius at $z=0$ over time. 
We don't show satellite galaxies on this plot out of an abundance of caution that their size evolution may be impacted by their infall to their parent halo in a way that is hard to disentangle.  However, the satellites tend to have the same age gradient trend as the field dwarfs, which we discuss in more detail in Section \ref{sec:environment_discussion}. 

To examine growth, we divide the sample into two, classifying each galaxy as having either a "steep" or "flat" age gradient. A steep (negative) gradient is any galaxy with a $dt_{90}/f_{R_e}$ value less than the average value of the sample, $-2.05 \text{ Gyr}$, while a flat gradient is any galaxy with $-2.05 \text{ Gyr} \leq dt_{90}/f_{R_e} \leq 0.25 \text{ Gyr}$. Thus, the magenta lines in Fig.~\ref{fig:sf_history} show the average $\text{SFR}_{90}/R_{e, z=0}$ or $R_e/R_{e, z=0}$ for all galaxies with stellar mass greater than $10^{5.8}$ $\Msol$ and a flat age gradient, while the blue lines show the average $y$-axis value for galaxies with a steep age gradient. We further divide the sample of flat gradients by mass, since Figs.~\ref{fig:dt90_t90_overview} and ~\ref{fig:core_dt90} both show that flat gradients exist at both high mass, cored galaxies and low mass, cuspy galaxies. The mass cutoff is at $10^{5.8}$ M$_\odot$. Thus, the yellow lines show the average effective radius and $\text{SFR}_{90}$ for field galaxies with flat age gradients and masses $< 10^{5.8}$ M$_{\odot}$. The shaded regions enclose 68\% of the sample. 


Overall, the right panel of Figure \ref{fig:sf_history} shows that the effective radii generally grow over time for all different classes of galaxies, including the flat, low mass field dwarfs.  The low mass field dwarfs with flat age gradients (predictably) have a decreasing $\text{SFR}_{90}$ early in their life (left panel), and past $\sim$4 Gyr the radius enclosing 90\% of the stars is 0 due to early quenching. However, their effective radius continues to increase even after they quench at $\sim4$Gyr and it is unclear if this is numerical or physical (e.g., due to mergers, shocks, etc).  We further discuss the possible role of resolution in artificially inflating the stars in these dwarfs in Section \ref{sec:res}.

Both panels of Fig.~\ref{fig:sf_history} show that the high mass field galaxies with a flat gradient grow more slowly at early times, but have more growth at recent times. In the left panel, the normalized $\text{SFR}_{90}$ increases with time, but starts off smaller than for the galaxies with steep gradients, before overtaking them and being noticeably larger than the steep gradients at $\sim$12-13.8 Gyr after the Big Bang. Overall, steep gradient field galaxies seem to experience a smaller change in the extent of their star formation compared to the galaxies with flatter gradients, i.e., the $\text{SFR}_{90}$ increases less than the flat gradient's $\text{SFR}_{90}$ value over time.  Furthermore, the spread of $\text{SFR}_{90}$ for field galaxies with steep gradients tends to be smaller than the more massive field galaxies with flat gradients. These trends confirm the growth patterns identified based on the example galaxies in Fig.~\ref{fig:r_age_plots}.

Switching our attention to the normalized effective radii with time for field galaxies (Fig.~\ref{fig:sf_history}, right panel), we notice that field galaxies with steep gradients seem to grow faster at early times, but the slope in their growth decreases at $\sim$8 Gyr.  Meanwhile, the field galaxies with flat age gradients at $z=0$ grow steadily in time, overtaking the galaxies with steep gradients at recent times. 



We have verified that the higher mass field dwarfs with flat age gradients have young stars that are forming not just on the outskirts of the galaxy, but at all radii.  
This fact can be seen in the age fraction for Elektra 2 in Figure \ref{fig:r_age_plots} as well. This distribution of star formation at later times flattens the age gradient by adding young stars across the galaxy, out to the radius to which the older stars have migrated. 

\subsection{Mergers}\label{sec:merger_discussion}



\begin{figure*}
    \centering
    \includegraphics[width=0.85\textwidth]{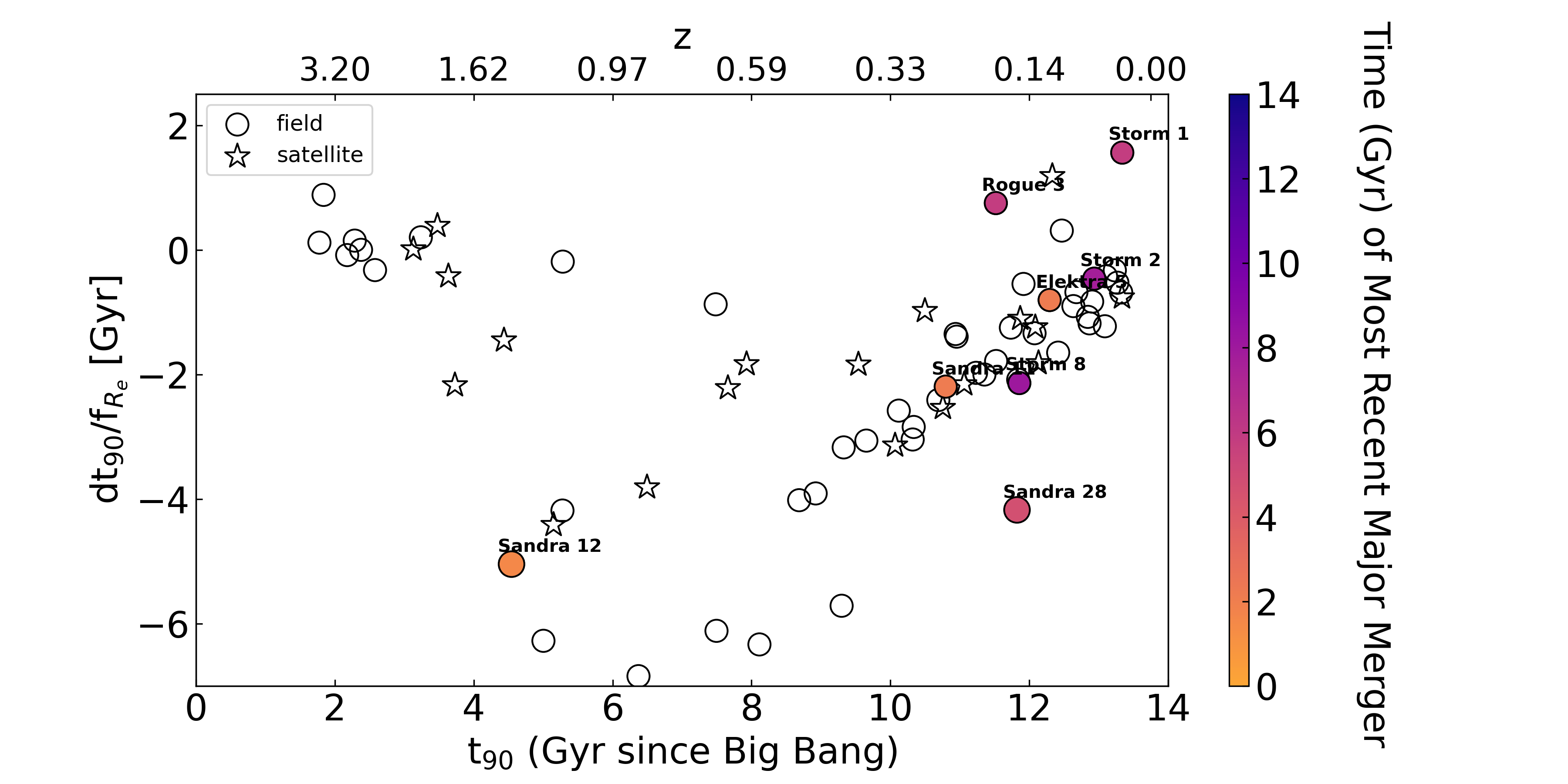}
    \caption{A recreation of Fig. \ref{fig:dt90_t90_overview}, but only the halos with a major merger are colored and identified. The colors correspond to the time after the Big Bang of their most recent major merger. }
    \label{fig:dt90_merger}
\end{figure*}

\begin{figure*}
    \centering
    \includegraphics[width=0.85\textwidth]{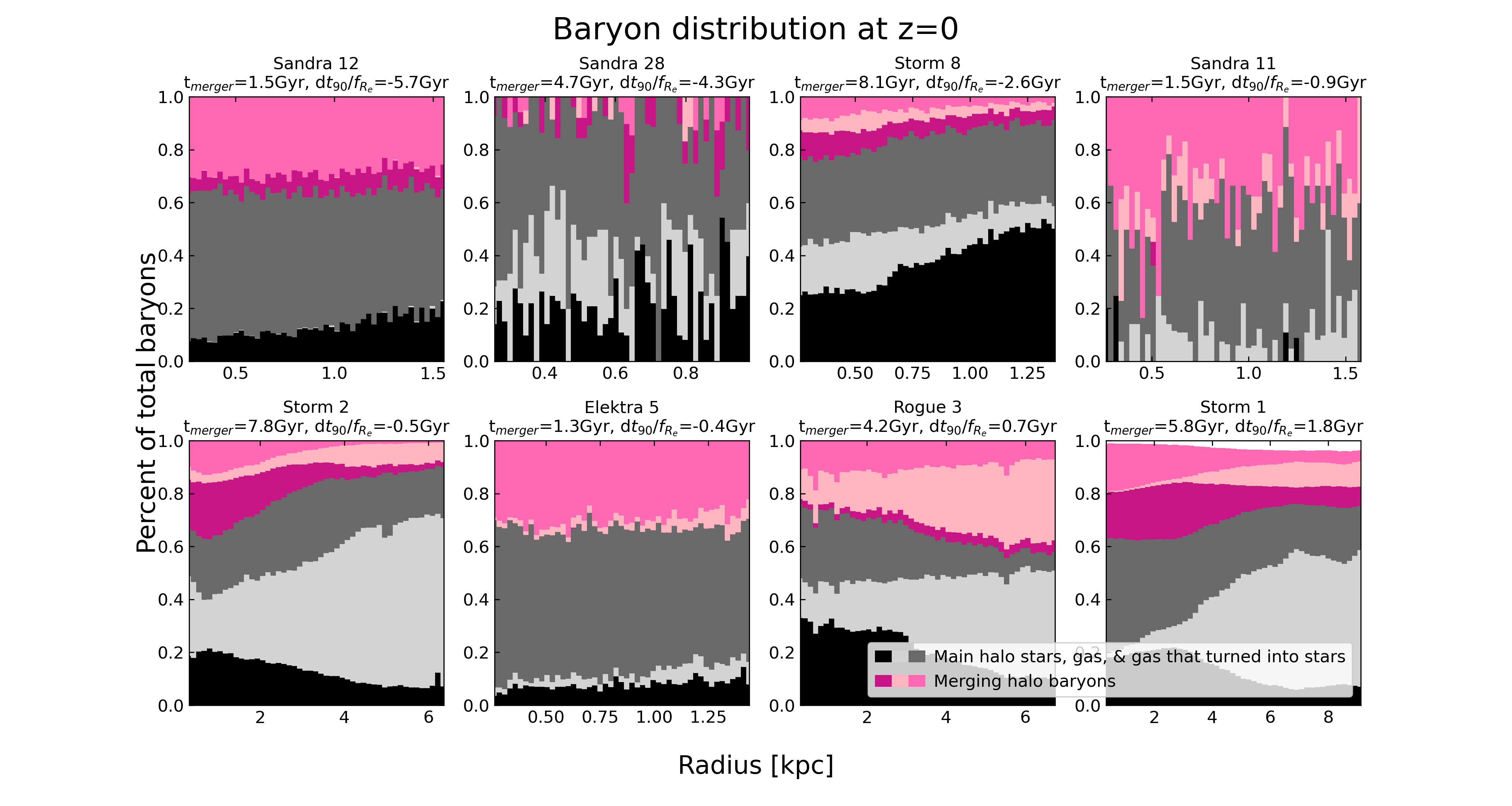}
    \caption{Radial distribution of gas and stars of each halo with a major merger in its history. 
    The black, dark gray, and light gray colors represent star, gas that turned into stars, and gas particles of the primary halo involved in the merger, and the varying shades of pink represent the same from the secondary halo involved in the merger. Note that Storm halo 1 had more than one merging halo. We show the contribution from the secondary only; the white space represents contributions from a smaller, tertiary halo.} 
    \label{fig:merger_baryons}
\end{figure*}

As discussed in Sec.~\ref{sec:intro}, some previous simulations predict that mergers can both steepen and flatten dwarf stellar age gradients \citep{BenitezLlambay2016, Graus_2019}. We investigate the merger history of our dwarfs to see if major mergers affect the age gradients. We find that 8 halos in our sample experience a major merger sometime in their formation history. We define a major merger as an event where the sum of the smaller halo masses involved in the merger is at least 25\% of the largest involved halo's mass. Fig.~\ref{fig:dt90_merger} recreates Fig.~\ref{fig:dt90_t90_overview} while only marking those halos that experience a major merger according to the time of their most recent major merger. We find that the mergers occur only in field galaxies, and exist across a range of stellar age gradients and stellar masses. For example, Storm 8, Sandra 12, and Sandra 28 all have steep gradients, whereas Elektra 5, Storm 1, Storm 2, Rogue 3, and Sandra 11 all have relatively flat gradients. Furthermore, it doesn't seem like the time of the merger impacts the gradient. For example, Sandra 12's merger occurred very early in its history (1.54 Gyr after the Big Bang), while Storm 8's merger occurred much later (8 Gyr after the Big Bang), yet they both have steep gradients. Meanwhile, Elektra 5's merger was also early (2.15 Gyr), but it has a rather flat age gradient.

We also examine where the baryons involved in mergers end up in the main galaxy at $z=0$ in Fig.~\ref{fig:merger_baryons}. We plot the stars, gas, and stars that were previously gas particles before the merger.  The grey scale shows these baryons for the main progenitor halo, and pink colors represent contributions from the smaller halos involved in the merger. (Storm 1 had two simultaneous mergers.  We show the contribution for the more massive contributor only.) First, we find that there is no trend to where the accreted baryons end up.  For example, in Storm 2 they end up more centrally concentrated, but in Rogue 3 the accreted baryons tend toward the outer parts of the galaxy.  In others, the distribution is very uniform across radii (e.g., Elektra 5, Sandra 12).  Likewise, sometimes the gas that had been in the main progenitor seems to form centrally-located stars (e.g., Storm 2 and Rogue 3), but in general the gas that was already in the main galaxy tends to form stars fairly uniformly after the merger.  

We are not able to identify any trends between the $z=0$ age gradient and the distribution of baryons from the merger. For example, looking at Storm 2, one might expect that since a high concentration of stars from the merging halo are in the center, that Storm 2 would have a steep, negative age gradient. However, this is not the case. Storm 2 has a fairly flat age gradient of -0.49. Meanwhile, the accreted baryons in the dwarfs with steep age gradients (Sandra 12, Sandra 28, Storm 8) show no uniformity in where the accreted baryons end up, or where star formation is induced in the main galaxy. Thus, it is hard to determine what affect, if any, mergers have on a galaxy's age gradient.

Overall, we are unable to find any meaningful correlation between stellar age gradients and the presence of a major merger in its history \citep[see also][]{Graus_2019}.  This result seems to be at odds with those in \citet{BenitezLlambay2016}, who found that mergers can create steep age (and metallicity) gradients in dSphs.   In their model, mergers kick old, metal-poor stars to large radii, while bringing in new gas that leads to younger, more metal-rich, centrally-concentrated star formation.  We do not rule out that this could happen in our simulations, 
but it seems that not all mergers leads to steep age gradients. We do not find any major mergers in the history of our simulated satellites galaxies, and only Storm 8 seems to show that pre-existing stars in the main progenitor end up dominating the outskirts.  We conclude that mergers do not play a dominating role in setting age gradients in our simulated dwarfs.

\subsection{$t_{50}$ Gradients}\label{sec:t50_results}
\begin{figure*}
    \centering
    \includegraphics[width=0.85\textwidth]{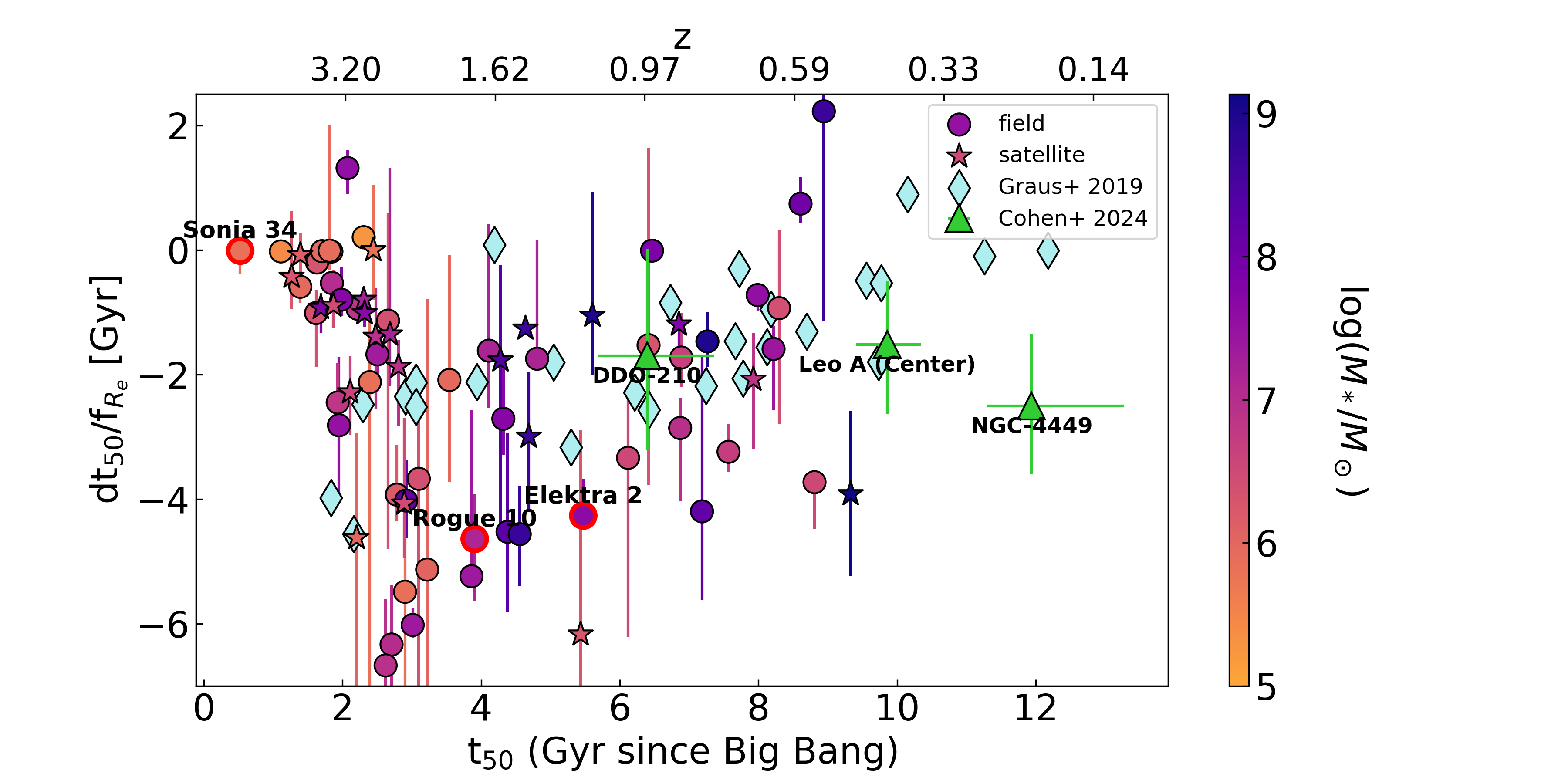}
    \caption{Here, we recreate Fig.~\ref{fig:dt90_t90_overview}, but use $t_{50}$ instead of using $t_{90}$. We show the age gradient $dt_{50}/f_{R_e}$ vs.~global $t_{50}$ in Gyr since Big Bang. We calculate $dt_{50}/f_{R_e}$ as we did for $dt_{90}/f_{R_e}$, now using the time when the galaxy or annulus formed 50\% of its stellar mass. Each point represents a simulated galaxy, with round points representing field dwarfs and stars representing satellite galaxies, with colors corresponding to the stellar mass of the galaxy. Light blue diamonds correspond to the galaxies presented in \protect\cite{Graus_2019} and green triangles correspond to a subset of galaxies presented in (R.~E.~Cohen et al., in prep). Unlike the data in Figure \ref{fig:dt90_t90_overview}, we do not find a strong correlation in $dt_{50}/f_{R_e}$ age gradient with global $t_{50}$.  This is in contrast to \citet{Graus_2019}, who find a strong age gradient trend using both $t_{50}$ and $t_{90}$. We infer this to mean that the age gradients in our simulated galaxies are established at later times (lower $z$) than those in \citet{Graus_2019}.}
    \label{fig:dt50_all}
\end{figure*}

\citet{Graus_2019} examined age gradients using both $dt_{90}/f_{R_e}$ and $dt_{50}/f_{R_e}$ for {\sc fire-2} galaxies, and found similar trends (i.e., slope of the relation between age gradient and global $t_{x0}$) in $t_{90}$ and $t_{50}$, respectively. Here we look at $dt_{50}$ values in our simulated galaxies. To calculate $dt_{50}/f_{R_e}$ we use the same method as was used to calculate $dt_{90}/f_{R_e}$ (see \ref{sec:gradients_def}), but instead determining the time at which 50\% of the stars have formed. The results are shown in Figure \ref{fig:dt50_all}.  As in Fig.~\ref{fig:dt90_t90_overview}, field (isolated) galaxies are shown as circles and satellite galaxies as stars, and the  color bar shows log stellar mass in units of solar mass.  We plot the global $t_{50}$ value of the galaxy in Gyr since the Big Bang on the $x$-axis and the stellar age gradient, $dt_{50}/f_{R_e}$, on the $y$-axis.\footnote{We do see that the cluster of halos with low stellar mass and early star formation have flat age gradients at both $t_{90}$ and $t_{50}$, further reaffirming that these halos have a flat gradient because they are quenched and therefore do not have a variety of stars within them.}  Again, we show the results from simulated galaxies in \citet{Graus_2019} as light blue diamonds, and results from preliminary observational analysis (Cohen et al., in preparation) as green triangles.

Unlike \citet{Graus_2019}, we do not find a similar trend in $dt_{50}/f_{R_e}$ as in $dt_{90}/f_{R_e}$. In fact, we find no evidence for a trend in $dt_{50}/f_{R_e}$ vs.~$t_{50}$ at all, despite our strong trend in $dt_{90}/f_{R_e}$ that was similar to what \citet{Graus_2019} found. We calculate the Pearson correlation coefficient and p-value for $dt_{50}/f_{R_e}$ vs. $t_{50}$ of both samples in order to quantify their differences. For \citet{Graus_2019}, we find a correlation coefficient $c_P = 0.73$ and a p-value of $2.47\times10^{-5}$, indicating that the values show a strong linear trend. For our sample of dwarfs, we find a correlation coefficient $c_P = -0.05$ and a p-value of $0.67$, indicating that there is no strong trend for our simulated dwarfs. It is plausible that the high p-value in our sample is due to the cluster of low-mass, quenched dwarfs, so we repeat the calculation for our sample after removing objects with $R_e < 450\text{ pc}$ (see Section \ref{sec:res}), which effectively removes all but three of these low-mass, quenched galaxies from our sample and makes our sample more similar to that in \citet{Graus_2019} in terms of stellar mass. Applying this cut, we find a correlation coefficient $c_P = -0.41$ and p-value $0.14$. The increased $c_P$ value indicates that there is slightly more of a trend between $dt_{50}$ and $t_{50}$ when removing the cluster of quenched dwarfs, but this value is still much lower than Graus' $c_{P}$. Thus, the age gradients at $t_{50}$  in the \citet{Graus_2019} sample have a close to linear trend with $t_{50}$, whereas our galaxies' $dt_{50}/f_{R_e}$ values do not correlate much (if at all) with $t_{50}$. From this difference, we infer that the process setting our age gradients (described in detail in the previous sections) is occurring later than the process that sets age gradients in {\sc fire-2}.  We discuss this further in Section \ref{sec:graus19_discussion}. 

Finally, we note that both the simulated dwarfs in \citet{Graus_2019} and the observed galaxies (Cohen et al., in preparation) extend to later values of $t_{50}$ than we see in our simulated sample.  With respect to the observations, the discrepancy could be due to the fact that we include all stars within the simulated galaxy to calculate $t_{50}$, a privilege that the observations do not have.  While the galaxies in the observational sample are chosen to have data that extends beyond $R_e$, the observations will still be biased against the oldest stars that are predicted to be in the outskirts by our simulated dwarfs \citep[see discussion in][]{Graus_2019}.  Meanwhile, \citet{Graus_2019} use all of the stars within 10\% of $R_{\rm vir}$ in their simulated dwarfs to calculate global $t_{50}$.  Again, this is likely to cut out older stars.  Overall, the radial extent to which stars are used to calculate global age gradients will bias our simulations to older values.

\section{Discussion}\label{sec:discussion}
Here, we discuss the implications of our findings and compare to previous work. 

\subsection{Resolution}\label{sec:res}

In Section \ref{sec:sfh} we noted that even the low mass, quenched field galaxies seem to experience reshuffling.  While we conclude that fluctuations in the gravitational potential well are responsible for reshuffling in more massive galaxies, this obviously cannot be the case in the quenched galaxies. They have not had an extended star formation history capable of creating dark matter cores, as evidenced in Figure \ref{fig:core_dt90}. It is possible that artificial two body relaxation is at play in this reshuffling.  \citet{Ludlow2019} show that interactions between two different mass particle species can lead to artificial inflation of stellar sizes in simulations.  The ratio between dark matter particle mass and the stellar particle mass is different in the two runs, but in both cases the dark matter particle masses are higher (15.8$\times$ higher in the Marvel runs, and 5.25$\times$ higher in the DCJL runs).  \citet{Ludlow2019} evaluates convergence criteria for sizes, suggesting that sizes could inflate up to $0.05 \times L$, where $L$ is the mean inter-particle separation.  A naive calculation based on the resolution of these simulations suggests that sizes in Marvel could be inflated up to $\sim$0.3 kpc, while sizes in DCJL could be inflated up to $\sim$0.8 kpc.  Hence, it is entirely possible that the increase in radius of the oldest stars is influenced by numerical noise. We note, however, that none of these effects can explain the large migrations seen in the more massive and (by definition) better resolved galaxies.  The distances migrated by the stars in these galaxies is well beyond any change explained by numerical effects.  We demonstrate in Appendix \ref{AppB} that the same process that creates dark matter cores is responsible for reshuffling the oldest stars in the more massive dwarfs.

In our analysis, we have excluded simulated galaxies with $R_e < 0.25$ kpc.  Although 0.25 kpc is considered resolved using the \citet{Power_2003} criteria, the \citet{Ludlow2020} criteria considers distances less than 0.45 kpc unresolved at these resolutions. 
However, we confirm that even if we remove the dwarfs with $R_e < 0.45$ kpc from Marvel and $R_e < 0.8$ kpc from DCJL, the U-shaped trend with gradients found in Fig.~\ref{fig:dt90_t90_overview} is still seen, suggesting the trend toward flatter gradients with decreasing $t_{90}$ is robust to resolution.  Likewise, we have confirmed that most of the quenched galaxies have stars that formed at similar radii over time.  Thurs, even if the reshuffling in these low mass dwarfs is numerical, they would still show flat gradients even without any reshuffling.  Finally, observationally the derivation of ages via color-magnitude diagram fitting encounters larger errors at old ages, suggesting that, to within observational errors, our prediction of a flat age gradient in quenched galaxies is robust. 

\subsection{Environment}\label{sec:environment_discussion}
Throughout our analysis, we have distinguished between field galaxies and satellite galaxies in order to identify any noticeable differences between the two samples. In most figures, satellites are denoted with star-shaped markers.  In the left panel of Fig.~\ref{fig:core_dt90}, we see that satellites follow the same mass trend as field galaxies, i.e., that more massive halos tend to have a flatter, more cored dark matter profiles, and vice versa for less massive halos.  The satellites with $M_* > 10^8$ M$_{\odot}$ are cored, but are also those with more recent $t_{90}$, as seen in the right panel of Fig.~\ref{fig:core_dt90}.  This is consistent with the results in \citet{Akins2021}, who used the DCJL simulations to study satellite quenching around Milky Way-mass galaxies and found that satellites with $M_* > 10^8$ M$_{\odot}$ tend not to quench by $z=0$.  In the $10^6 < M_*/M_{\odot} < 10^8$ mass range, \citet{Akins2021} found that quenching occurred very rapidly ($\lesssim$ 2 Gyr) after infall.  They found a mix of quenched and unquenched satellites, and a strong correlation with infall times, such that the unquenched galaxies in this mass range fell into their parent halo $<$1 Gyr before $z=0$.  Likewise, \citet{Engler2023} examines the SFHs, $t_{50}$, and $t_{90}$ values for satellites of Milky Way-mass galaxies in the TNG50 simulations. They find that the times of satellites' first infall correlate with their gas fractions, such that satellites whose first infall occurred more than 2.5 Gyr ago have extremely low gas fractions. They find that this process is even more efficient for dwarfs with lower masses. Consistent with this, we find that most of the satellites with $M_* < 10^8$ M$_{\odot}$ have $t_{90} < 8$ Gyr (as shown by the unfilled symbols in the right panel of Fig.~\ref{fig:core_dt90}). We also examined whether the time of gas stripping in satellites correlated with any trends in gradients or $t_{90}$, but were unable to identify any. Likewise, it is not clear what impact stripping of satellites might play in our results, since stripping can influence both gradients and $t_{90}$, such that satellites might simply evolve along the trend we find, or might move off of it. A deeper examination of any role of gas stripping on gradients would benefit from controlled experiments beyond the scope of this paper.

Overall, we find that there are no distinguishable age gradient trends in the satellites compared to field dwarfs.  Perhaps because the more massive satellites fell in recently and are able to maintain star formation after infall, the cored ($\alpha > -0.5$) satellite galaxies apparently follow similar trends to the field dwarfs, i.e., those with the most recent $t_{90}$ have the flattest age gradients, while those with $t_{90} \sim 8$ Gyr have steeper age gradients.  The satellite galaxies with $M_* < 10^8$ M$_{\odot}$ instead have age gradients that are most strongly correlated with dark matter core slope.  Those with the lowest masses have the steepest dark matter density profile slopes, along with the flattest age gradients.  As mass increase up to  $M_* = 10^8$, the density slope flattens and age gradients becomes steeper, as in the field dwarfs.  
We note that these trends are also consistent with the results presented in \citet{Read2019}, in which it is inferred that dSph galaxies with extended star formation, $t_{90} > 8$ Gyr, have cored dark matter density profiles based on their central densities.  We have two simulated dSphs that are cored with $t_{90} < 8$ Gyr, but they are also relatively massive \citep[$M_* > 10^7$ M$_{\odot}$, a mass range predicted to produce dark matter cores in our simulations;][]{Azartash-Namin2024}.  The results in \citet{Read2019} are also consistent with all of the observed dwarfs in this mass range having dark matter cores. 

The result that both simulated satellites and field dwarfs commonly show ``outside-in'' age gradients is consistent with observations.  As discussed in Section~\ref{sec:intro}, dwarfs of all types show outside-in age gradients.  This consistency across environments motivates a desire to find a physical mechanism that sets age gradients independent of environment.  In this work we find that age gradients are first dependent on mass \citep[and hence the ability to drive fluctuations in the gravitational potential well, e.g.,][]{Governato2012, DiCintio_2014, Tollet2016, Lazar_2020, Jackson2024}, with a secondary dependence on the duration of star formation for the most massive dwarfs.  While environment certainly has the ability to impact the duration of star formation \citep[e.g.,][]{Simpson2018, Akins2021, Joshi2021, Christensen2024}, it seems it plays a tertiary role in setting age gradients.

\subsection{Comparison with Graus et al. 2019 }\label{sec:graus19_discussion}
We compare our results directly with \cite{Graus_2019} in Figs. \ref{fig:dt90_t90_overview} and \ref{fig:dt50_all}. 
Although we found that our $t_{90}$ age gradients at $z=0$ match those of \cite{Graus_2019}, our age gradients at $t_{50}$ had very different results compared to \cite{Graus_2019}. The Pearson correlation coefficient confirms that \cite{Graus_2019} has a strong linear trend at $z=0$ between $dt_{50}/f_{R_e}$ and global $t_{50}$.  On the other hand, we find a much weaker correlation, if any. 

Similar to what we have found, \citet{Graus_2019} finds a correlation between age gradients and central dark matter density, as well as a trend for flatter age gradients to occur in those galaxies with young star formation that extended out to larger radii at late times.  These results describe those found in this work with $t_{90}$, yet we do not see that these processes have imprinted on $t_{50}$. This suggests that the processes that set age gradient are happening earlier in {\sc fire} than in our simulations, i.e., prior to $t_{50}$. 

We infer that the differences in our $t_{50}$ analyses arise from differences in how feedback is implemented in our two simulations. 
\citet{ElBadry_2016} showed that simulated {\sc fire} galaxies that undergo significant potential well fluctuations due to bursty feedback also undergo inflows and outflows of star forming gas, i.e., the ``breathing'' process described in Section~\ref{sec:breathing}.  Their results suggest that the breathing mode accompanies gravitational potential well fluctuations in {\sc fire}.  However, while we form dark matter cores through fluctuations in the gravitational potential, we find no strong evidence for an accompanying ``breathing mode,'' as evidenced in Fig.~\ref{fig:radial_migration}.  \citet{ElBadry_2016} demonstrated that the {\sc fire} dwarfs undergo significant radial velocity oscillations in their star forming gas and young stars (at least $\pm$20 km/s) within the first few Gyr after the Big Bang, but no such oscillations are evident in our simulated dwarfs. Thus, we tentatively conclude that fluctuations in the gravitational potential well are decoupled from a breathing mode in our simulations, though we will investigate this further in future work. 

The existence of the breathing mode in {\sc fire} may explain why negative age gradients form more quickly in their simulated dwarfs relative to ours. The breathing mode provides a mechanism that reshuffles the older stars earlier in {\sc fire} than is happening in our simulations.  They are thus able to create an age gradient early, imprinting on $t_{50}$.  On the other hand, our lack of breathing mode at early times delays reshuffling of the older stars, so that age gradients are most strongly imposed sometime after $t_{50}$, and hence only appear strongly in $t_{90}$.  This is consistent with the idea that stronger cores are developed with sustained star formation \citep[e.g.,][]{Read2019, Muni2024}.  In this scenario, longer star formation leads to more sustained gravitational potential well fluctuations, and hence longer periods of time in which old stars can be kicked out from the central regions to the outskirts.

In summary, age gradients offer another mechanism for testing feedback models in cosmological simulations.  \citet{ElBadry_2016} demonstrated that the breathing mode in {\sc fire} galaxies leads to a predicted correlation between sSFR and galaxy size, or sSFR and velocity dispersion.  These predictions have been put to the test, but with mixed results \citep{Patel2018, Patel2023, Hirtenstein2019, Pelliccia2020, Emami2021}, and the {\sc fire} bursts seem too short, too intense, and to have a different duty cycle than observations \citep[e.g.,][]{McQuinn2010}. While we intend to explore these trends further in our simulations in future work, here we have shown that resolved stellar populations offer another method for testing these models and constraining the strength/impact of feedback, by measuring $t_{50}$ as a function of radius.  Measurements of the age gradients in dwarf galaxies are already underway (Cohen et al.~2024, in prep.) using archival Hubble Space Telescope resolved stellar population observations, and these observations may be able to test the various model predictions, assuming the data is deep enough and extends to sufficient radii to recover $t_{50}$ gradients.  

\subsection{What About Metallicity Gradients?}

In this work, we focus on age gradients.  Observationally, a determination of age gradients is time-consuming to quantify, requiring significant data, e.g., either detailed resolved stellar populations as a function of radius, or multi-wavelength photometric data that allows an inference of the stellar mass and star formation history as a function of age.  On the other hand, metallicity gradients in dwarfs have been easier to achieve by targeting individual stars and obtaining spectra across the radii of galaxies. However, it is not immediately clear if the processes that set metallicity gradients are the same as those that set age gradients.

A recent uniform re-examination of metallicity gradients in Local Group dwarfs finds no correlation with metallicity gradients and various properties such as stellar mass, star formation timescales such at $t_{50}$ or $t_{90}$, kinematics, morphology, or environment \citep{Taibi2022}. \citet{Taibi2022} utilizes publicly available spectroscopic red giant star catalogs to compute the radial metallicity gradients of 30 Local Group dwarfs, the largest uniform determination to date.  The only correlation identified was that galaxies with the strongest metallicity gradients tended to be those with kinematic evidence for a past merger event, in agreement with \citet{BenitezLlambay2016}. They also collected metallicity info for stars in a number of simulated data sets, including the  Marvel and DCJL samples, and derived their metallicity gradients using the same algorithm applied to the observations.  Their figure 9 shows our simulated metallicity gradients compared to observations.  There is a reasonably good agreement between the simulations and the observations, with the exception that we do not produce the strongest metallicity gradients seen in the observed dwarfs, which are also the dwarfs with kinematic evidence for a past merger.  Interestingly, our simulations show a hint of a U-shaped trend as a function of stellar mass.  Although $t_{50}$ and $t_{90}$ are plotted in terms of the logarithmic lookback time, it can be seen in their figure 9 that there is a dip between 9.7 $<$ log($t_{90}$) $<$ 10.1, i.e, between lookback times of 5 to 12 Gyr.  In other words, we see a similar U-shape trend imprinted in metallicity as we do in age gradients.  However, we find no correlation between the age and metallicity gradients.  In other words, while both have a U-shape trend with $t_{90}$, the galaxies with strong age gradients are not necessarily those with the strongest metallicity gradients.    

\citet{Mercado2021} showed that the same processes that create age gradients in the {\sc fire}-2 dwarfs (strong reshuffling of the oldest stars, with a secondary dependency on the extent of low $z$ star formation) also create metallicity gradients, such that metallicity gradients in {\sc fire}-2 correlate with $t_{50}$, like age gradients. However, \citet{Taibi2022} showed that the strong correlation found in \citet{Mercado2021} appears to be at odds with a large sample of observed dwarf galaxies \citep[though see][which use photometric narrow band imaging allowing for higher completeness than spectroscopic surveys]{Fu2024a, Fu2024b}.  Additionally, \citet{Cardona-Barrero2023}, examine metallicity gradients using the NIHAO simulations and finds results consistent with \citet{Taibi2022}. They find that the strong metallicity gradients in \citet{Mercado2021} are only reproduced if rotationally-supported galaxies are excluded. In our sample, we have identified simulated galaxies that have a relatively flat metallicity gradient while having a strong outside-in age gradient.  Hence, it is not clear if it is the same processes creating metallicity gradients that also create age gradients, despite the similarity in U-shape trends with $t_{90}$.  We leave a full investigation to future work.

Meanwhile, only mergers can create strong age and metallicity gradients in the simulations of \citet{BenitezLlambay2016}, because their simulations do not induce gravitational potential well fluctuations with bursty feedback \citep{BenitezLlambay2019}.  Overall, these comparisons highlight that gradients offer a testable constraint on different feedback models in cosmological simulations.  

\section{Conclusions}\label{sec:conclusion}
Age gradients in dwarf galaxies make it appear as though dwarfs formed from the outside-in, with older stars often found at larger radii than younger stars. This is in contrast to the better understood inside-out formation appearance of Milky Way-type galaxies, which is often interpreted within the context of angular momentum acquisition in disk galaxy formation \citep[e.g.,][]{Barnes2014, Bird2021}. However, the outside-in gradients in dwarf galaxies have generally lacked a satisfactory explanation, particularly since such gradients seem to be independent of environment, and many explanations are environmentally-dependent.  In this paper, we examined a sample of 73 fully-cosmological, well-resolved, simulated dwarf galaxies in order to investigate where these outside-in gradients originate.  Our sample extends from $\sim$10$^6$ M$_{\odot}$ to 10$^9$ M$_{\odot}$ in stellar mass, and includes both isolated field dwarfs and dwarfs that are satellites of a larger system.  This is the largest and most diverse sample of simulated dwarfs used to study the origin of age gradients.  

We first measure age gradients using $t_{90}$, the time at which 90\% of the stellar mass has formed, and take the gradient to be the difference in $t_{90}$ between 1.5 $R_e$ and the center of the galaxy, normalized by $R_e$.  In agreement with previous works \citep{Graus_2019}, we find that dwarf galaxies with the most extended SFH histories have flat age gradients, but that the age gradients steepen (becoming more strongly outside-in) as $t_{90}$ becomes less recent. 
Our larger sample allows us to predict a turn-over for dwarfs with $t_{90}$ earlier than 8 Gyr: age gradients also become flatter as $t_{90}$ decreases.  Overall, this creates a ``U''-shaped prediction (see Figure \ref{fig:dt90_t90_overview}).

In investigating the origin of these gradients, we find that dwarf galaxies {\it do} generally form their stars from the inside-out (or otherwise with a similar radius over time, but not outside-in), just like more massive galaxies.  However, the oldest, most-centrally located stars are reshuffled, so that they end up at the largest radii.  
We find that there are two main processes that determine dwarf galaxy age gradients. First, the same process that creates dark matter cores (fluctuation of the gravitational potential well) also reshuffles the old stars. 
Low-mass halos with a cuspy (steeply rising) dark matter density profiles tend to have flat gradients, but age gradients steepen as the dark matter profile becomes more cored (left panel Fig.~\ref{fig:core_dt90}). However, the galaxies with the most cored dark matter profiles show a secondary effect, such that they can have a range of age gradients:  The more extended the SFH, the flatter the age gradient. 

We have verified (Figure \ref{fig:sf_history}) that the more massive dwarfs with flatter age gradients have had more recent size growth than their counterparts with steep age gradients.  The more recent size growth includes young stars at all radii, including the radii to which the older stars have been reshuffled.  This ``erases'' the steep gradient imparted by the gravitational potential well fluctuation, creating a flat age gradient.

These processes are generally independent of environment, which leads to no discernible difference in the age gradients of satellites compared to isolated, field dwarfs.  The ability of the galaxy to fluctuate its gravitational potential via bursty feedback is thought to be strongly dependent on mass \citep[e.g.,][]{Governato2012, DiCintio_2014, Tollet2016, Lazar_2020, Jackson2024, Azartash-Namin2024}, and it is yet unclear what sets the time at which galaxies grow in size.  Certainly we might expect environment to impact the duration of star formation.  However, the lowest-mass satellites quench early, even before accretion onto their host halo, and the most massive satellites are hard to quench before $z=0$ and have recent infall times (because they are otherwise destroyed by dynamical friction quickly) \citep[e.g.,][]{Simpson2018, Akins2021, Joshi2021, Christensen2024}.  Because the duration of star formation depends on the mass of the satellite, the overall trend imparted by mass remains the dominant impact on the age gradients in the satellites, with environment playing a tertiary role.  The fact that outside-in age gradients are common in dwarfs found in all environments supports the idea that processes independent of environment, such as the ones identified here, must be the dominant factor in setting age gradients.  
We also explore the role of major mergers in setting age gradients, but found no conclusive impact.  

While our simulations and those run with the {\sc fire}-2 model both reshuffle stars while inducing dark matter cores, the timescales over which the process occurs seems to be quite different.  
While \citet{Graus_2019} find similar age gradients in $t_{50}$ and $t_{90}$, we do not.  We find no correlation with $dt_{50}/f_{R_e}$ and $t_{50}$ (see Figure \ref{fig:dt50_all}).  Thus, the reshuffling process sets age gradients by $t_{50}$ in {\sc fire}-2, while it occurs later in our simulations, sometime after $t_{50}$.  We speculate that this is due to the early ``breathing mode'' shown in \citet{ElBadry_2016}, in which star forming gas experiences radial inflows and outflows on the order of tens of km/s, but does not seem happen in our simulations (Figure \ref{fig:radial_migration}).  Hence, resolved stellar population data that allows to probe $t_{50}$ and $dt_{50}/f_{R_e}$ has the potential to constrain these different feedback effects.  

\begin{acknowledgments}
AMB and CLR thank Kristy McQuinn and Salvatore Taibi for useful conversations that improved this draft, and Jordan Van Nest for providing the simulation slope measurements used in Figure \ref{fig:core_dt90}. AMB and CLR acknowledge support from HST AR-17550 provided by NASA through a grant from the Space Telescope Science Institute, which is operated by the Association of Universities for Research in Astronomy, Incorporated, under NASA contract NAS5-26555.  FDM and AMB acknowledge support from HST AR-13925. AMB acknowledges support from HST AR-14281. REC acknowledges support from HST AR-17038. CRC acknowledges support from the NSF under CAREER grant No. AST-1848107.  Resources supporting this work were provided by the NASA High-End Computing (HEC) Program through the NASA Advanced Supercomputing (NAS) Division at Ames Research Center.  FDM is grateful for the hospitality of Perimeter Institute where part of this work was carried out. Research at Perimeter Institute is supported in part by the Government of Canada through the Department of Innovation, Science and Economic Development and by the Province of Ontario through the Ministry of Colleges and Universities. This research was also supported in part by the Simons Foundation through the Simons Foundation Emmy Noether Fellows Program at Perimeter Institute. FDM acknowledges support from NSF grant PHY2013909. This research was supported in part by grant NSF PHY-2309135 to the Kavli Institute for Theoretical Physics (KITP).
\end{acknowledgments}

%

\vspace{5mm}

\bibliography{gradients}{}

\begin{thebibliography}{}
\expandafter\ifx\csname natexlab\endcsname\relax\def\natexlab#1{#1}\fi
\providecommand{\url}[1]{\href{#1}{#1}}
\providecommand{\dodoi}[1]{doi:~\href{http://doi.org/#1}{\nolinkurl{#1}}}
\providecommand{\doeprint}[1]{\href{http://ascl.net/#1}{\nolinkurl{http://ascl.net/#1}}}
\providecommand{\doarXiv}[1]{\href{https://arxiv.org/abs/#1}{\nolinkurl{https://arxiv.org/abs/#1}}}

\bibitem[{{Agertz} {et~al.}(2013){Agertz}, {Kravtsov}, {Leitner}, \&
  {Gnedin}}]{Agertz2013}
{Agertz}, O., {Kravtsov}, A.~V., {Leitner}, S.~N., \& {Gnedin}, N.~Y. 2013,
  \apj, 770, 25, \dodoi{10.1088/0004-637X/770/1/25}

\bibitem[{{Agertz} {et~al.}(2021){Agertz}, {Renaud}, {Feltzing}, {Read},
  {Ryde}, {Andersson}, {Rey}, {Bensby}, \& {Feuillet}}]{Agertz2021}
{Agertz}, O., {Renaud}, F., {Feltzing}, S., {et~al.} 2021, \mnras, 503, 5826,
  \dodoi{10.1093/mnras/stab322}

\bibitem[{{Akins} {et~al.}(2021){Akins}, {Christensen}, {Brooks}, {Munshi},
  {Applebaum}, {Engelhardt}, \& {Chamberland}}]{Akins2021}
{Akins}, H.~B., {Christensen}, C.~R., {Brooks}, A.~M., {et~al.} 2021, \apj,
  909, 139, \dodoi{10.3847/1538-4357/abe2ab}

\bibitem[{{Albers} {et~al.}(2019){Albers}, {Weisz}, {Cole}, {Dolphin},
  {Skillman}, {Williams}, {Boylan-Kolchin}, {Bullock}, {Dalcanton}, {Hopkins},
  {Leaman}, {McConnachie}, {Vogelsberger}, \& {Wetzel}}]{Albers2019}
{Albers}, S.~M., {Weisz}, D.~R., {Cole}, A.~A., {et~al.} 2019, \mnras, 490,
  5538, \dodoi{10.1093/mnras/stz2903}

\bibitem[{{Aparicio} \& {Tikhonov}(2000)}]{Aparicio2000}
{Aparicio}, A., \& {Tikhonov}, N. 2000, \aj, 119, 2183, \dodoi{10.1086/301360}

\bibitem[{{Azartash-Namin} {et~al.}(2024{\natexlab{a}}){Azartash-Namin},
  {Engelhardt}, {Munshi}, {Keller}, {Brooks}, {Van Nest}, {Christensen},
  {Quinn}, \& {Wadsley}}]{Azartash2024}
{Azartash-Namin}, B., {Engelhardt}, A., {Munshi}, F., {et~al.}
  2024{\natexlab{a}}, arXiv e-prints, arXiv:2401.06041,
  \dodoi{10.48550/arXiv.2401.06041}

\bibitem[{{Azartash-Namin} {et~al.}(2024{\natexlab{b}}){Azartash-Namin},
  {Engelhardt}, {Munshi}, {Keller}, {Brooks}, {Van Nest}, {Christensen},
  {Quinn}, \& {Wadsley}}]{Azartash-Namin2024}
---. 2024{\natexlab{b}}, arXiv e-prints, arXiv:2401.06041,
  \dodoi{10.48550/arXiv.2401.06041}

\bibitem[{{Barnes} {et~al.}(2014){Barnes}, {van Zee}, {Dale}, {Staudaher},
  {Bullock}, {Calzetti}, {Chandar}, \& {Dalcanton}}]{Barnes2014}
{Barnes}, K.~L., {van Zee}, L., {Dale}, D.~A., {et~al.} 2014, \apj, 789, 126,
  \dodoi{10.1088/0004-637X/789/2/126}

\bibitem[{{Battaglia} {et~al.}(2008){Battaglia}, {Helmi}, {Tolstoy}, {Irwin},
  {Hill}, \& {Jablonka}}]{Battaglia2008}
{Battaglia}, G., {Helmi}, A., {Tolstoy}, E., {et~al.} 2008, \apjl, 681, L13,
  \dodoi{10.1086/590179}

\bibitem[{{Battaglia} {et~al.}(2011){Battaglia}, {Tolstoy}, {Helmi}, {Irwin},
  {Parisi}, {Hill}, \& {Jablonka}}]{Battaglia2011}
{Battaglia}, G., {Tolstoy}, E., {Helmi}, A., {et~al.} 2011, \mnras, 411, 1013,
  \dodoi{10.1111/j.1365-2966.2010.17745.x}

\bibitem[{{Battaglia} {et~al.}(2006){Battaglia}, {Tolstoy}, {Helmi}, {Irwin},
  {Letarte}, {Jablonka}, {Hill}, {Venn}, {Shetrone}, {Arimoto}, {Primas},
  {Kaufer}, {Francois}, {Szeifert}, {Abel}, \& {Sadakane}}]{Battaglia2006}
---. 2006, \aap, 459, 423, \dodoi{10.1051/0004-6361:20065720}

\bibitem[{{Bellazzini} {et~al.}(2014){Bellazzini}, {Beccari}, {Fraternali},
  {Oosterloo}, {Sollima}, {Testa}, {Galleti}, {Perina}, {Faccini}, \&
  {Cusano}}]{Bellazzini2014}
{Bellazzini}, M., {Beccari}, G., {Fraternali}, F., {et~al.} 2014, \aap, 566,
  A44, \dodoi{10.1051/0004-6361/201423659}

\bibitem[{{Bellovary} {et~al.}(2019){Bellovary}, {Cleary}, {Munshi}, {Tremmel},
  {Christensen}, {Brooks}, \& {Quinn}}]{Bellovary_2019}
{Bellovary}, J.~M., {Cleary}, C.~E., {Munshi}, F., {et~al.} 2019, \mnras, 482,
  2913, \dodoi{10.1093/mnras/sty2842}

\bibitem[{{Ben{\'\i}tez-Llambay} {et~al.}(2019){Ben{\'\i}tez-Llambay}, {Frenk},
  {Ludlow}, \& {Navarro}}]{BenitezLlambay2019}
{Ben{\'\i}tez-Llambay}, A., {Frenk}, C.~S., {Ludlow}, A.~D., \& {Navarro},
  J.~F. 2019, \mnras, 488, 2387, \dodoi{10.1093/mnras/stz1890}

\bibitem[{{Ben{\'\i}tez-Llambay} {et~al.}(2016){Ben{\'\i}tez-Llambay},
  {Navarro}, {Abadi}, {Gottl{\"o}ber}, {Yepes}, {Hoffman}, \&
  {Steinmetz}}]{BenitezLlambay2016}
{Ben{\'\i}tez-Llambay}, A., {Navarro}, J.~F., {Abadi}, M.~G., {et~al.} 2016,
  \mnras, 456, 1185, \dodoi{10.1093/mnras/stv2722}

\bibitem[{{Bettinelli} {et~al.}(2019){Bettinelli}, {Hidalgo}, {Cassisi},
  {Aparicio}, {Piotto}, {Valdes}, \& {Walker}}]{Bettinelli2019}
{Bettinelli}, M., {Hidalgo}, S.~L., {Cassisi}, S., {et~al.} 2019, \mnras, 487,
  5862, \dodoi{10.1093/mnras/stz1679}

\bibitem[{{Bird} {et~al.}(2021){Bird}, {Loebman}, {Weinberg}, {Brooks},
  {Quinn}, \& {Christensen}}]{Bird2021}
{Bird}, J.~C., {Loebman}, S.~R., {Weinberg}, D.~H., {et~al.} 2021, \mnras, 503,
  1815, \dodoi{10.1093/mnras/stab289}

\bibitem[{{Brook} \& {Di Cintio}(2015)}]{Brook2015}
{Brook}, C.~B., \& {Di Cintio}, A. 2015, \mnras, 450, 3920,
  \dodoi{10.1093/mnras/stv864}

\bibitem[{{Brooks} {et~al.}(2007){Brooks}, {Governato}, {Booth}, {Willman},
  {Gardner}, {Wadsley}, {Stinson}, \& {Quinn}}]{Brooks2007}
{Brooks}, A.~M., {Governato}, F., {Booth}, C.~M., {et~al.} 2007, \apjl, 655,
  L17, \dodoi{10.1086/511765}

\bibitem[{{Brooks} \& {Zolotov}(2014)}]{Brooks2014}
{Brooks}, A.~M., \& {Zolotov}, A. 2014, \apj, 786, 87,
  \dodoi{10.1088/0004-637X/786/2/87}

\bibitem[{{Brooks} {et~al.}(2011){Brooks}, {Solomon}, {Governato}, {McCleary},
  {MacArthur}, {Brook}, {Jonsson}, {Quinn}, \& {Wadsley}}]{Brooks2011}
{Brooks}, A.~M., {Solomon}, A.~R., {Governato}, F., {et~al.} 2011, \apj, 728,
  51, \dodoi{10.1088/0004-637X/728/1/51}

\bibitem[{{Burger} {et~al.}(2022){Burger}, {Zavala}, {Sales}, {Vogelsberger},
  {Marinacci}, \& {Torrey}}]{Burger2022}
{Burger}, J.~D., {Zavala}, J., {Sales}, L.~V., {et~al.} 2022, \mnras, 513,
  3458, \dodoi{10.1093/mnras/stac994}

\bibitem[{{Cardona-Barrero} {et~al.}(2023){Cardona-Barrero}, {Di Cintio},
  {Battaglia}, {Macci{\`o}}, \& {Taibi}}]{Cardona-Barrero2023}
{Cardona-Barrero}, S., {Di Cintio}, A., {Battaglia}, G., {Macci{\`o}}, A.~V.,
  \& {Taibi}, S. 2023, \mnras, 519, 1545, \dodoi{10.1093/mnras/stac3243}

\bibitem[{Chan {et~al.}(2015)Chan, Kereš, Oñorbe, Hopkins, Muratov,
  Faucher-Giguère, \& Quataert}]{Chan_2015}
Chan, T.~K., Kereš, D., Oñorbe, J., {et~al.} 2015, Monthly Notices of the
  Royal Astronomical Society, 454, 2981–3001, \dodoi{10.1093/mnras/stv2165}

\bibitem[{Chiappini {et~al.}(2001)Chiappini, Matteucci, \&
  Romano}]{Chiappini_2001}
Chiappini, C., Matteucci, F., \& Romano, D. 2001, The Astrophysical Journal,
  554, 1044–1058, \dodoi{10.1086/321427}

\bibitem[{{Christensen} {et~al.}(2012){Christensen}, {Quinn}, {Governato},
  {Stilp}, {Shen}, \& {Wadsley}}]{Christensen_2012}
{Christensen}, C., {Quinn}, T., {Governato}, F., {et~al.} 2012, \mnras, 425,
  3058, \dodoi{10.1111/j.1365-2966.2012.21628.x}

\bibitem[{{Christensen} {et~al.}(2024){Christensen}, {Brooks}, {Munshi},
  {Riggs}, {Van Nest}, {Akins}, {Quinn}, \& {Chamberland}}]{Christensen2024}
{Christensen}, C.~R., {Brooks}, A.~M., {Munshi}, F., {et~al.} 2024, \apj, 961,
  236, \dodoi{10.3847/1538-4357/ad0c5a}

\bibitem[{{Christensen} {et~al.}(2016){Christensen}, {Dav{\'e}}, {Governato},
  {Pontzen}, {Brooks}, {Munshi}, {Quinn}, \& {Wadsley}}]{Christensen2016}
{Christensen}, C.~R., {Dav{\'e}}, R., {Governato}, F., {et~al.} 2016, \apj,
  824, 57, \dodoi{10.3847/0004-637X/824/1/57}

\bibitem[{{Dale} {et~al.}(2016){Dale}, {Beltz-Mohrmann}, {Egan}, {Hatlestad},
  {Herzog}, {Leung}, {McLane}, {Phenicie}, {Roberts}, {Barnes}, {Boquien},
  {Calzetti}, {Cook}, {Kobulnicky}, {Staudaher}, \& {van Zee}}]{Dale2016}
{Dale}, D.~A., {Beltz-Mohrmann}, G.~D., {Egan}, A.~A., {et~al.} 2016, \aj, 151,
  4, \dodoi{10.3847/0004-6256/151/1/4}

\bibitem[{{Daniel} \& {Wyse}(2015)}]{Daniel2015}
{Daniel}, K.~J., \& {Wyse}, R. F.~G. 2015, \mnras, 447, 3576,
  \dodoi{10.1093/mnras/stu2683}

\bibitem[{{Deason} {et~al.}(2014){Deason}, {Wetzel}, \&
  {Garrison-Kimmel}}]{Deason2014}
{Deason}, A., {Wetzel}, A., \& {Garrison-Kimmel}, S. 2014, \apj, 794, 115,
  \dodoi{10.1088/0004-637X/794/2/115}

\bibitem[{{Debattista} {et~al.}(2006){Debattista}, {Mayer}, {Carollo}, {Moore},
  {Wadsley}, \& {Quinn}}]{Debattista2006}
{Debattista}, V.~P., {Mayer}, L., {Carollo}, C.~M., {et~al.} 2006, \apj, 645,
  209, \dodoi{10.1086/504147}

\bibitem[{{del Pino} {et~al.}(2015){del Pino}, {Aparicio}, \&
  {Hidalgo}}]{delPino2015}
{del Pino}, A., {Aparicio}, A., \& {Hidalgo}, S.~L. 2015, \mnras, 454, 3996,
  \dodoi{10.1093/mnras/stv2174}

\bibitem[{DiCintio {et~al.}(2014)DiCintio, Brook, Dutton, Macciò, Stinson, \&
  Knebe}]{DiCintio_2014}
DiCintio, A., Brook, C.~B., Dutton, A.~A., {et~al.} 2014, Monthly Notices of
  the Royal Astronomical Society, 441, 2986–2995,
  \dodoi{10.1093/mnras/stu729}

\bibitem[{{Dutton} {et~al.}(2019){Dutton}, {Macci{\`o}}, {Buck}, {Dixon},
  {Blank}, \& {Obreja}}]{Dutton2019}
{Dutton}, A.~A., {Macci{\`o}}, A.~V., {Buck}, T., {et~al.} 2019, \mnras, 486,
  655, \dodoi{10.1093/mnras/stz889}

\bibitem[{El-Badry {et~al.}(2016)El-Badry, Wetzel, Geha, Hopkins, Kereš, Chan,
  \& Faucher-Giguère}]{ElBadry_2016}
El-Badry, K., Wetzel, A., Geha, M., {et~al.} 2016, The Astrophysical Journal,
  820, 131, \dodoi{10.3847/0004-637x/820/2/131}

\bibitem[{{Elmegreen} {et~al.}(2014){Elmegreen}, {Struck}, \&
  {Hunter}}]{Elmegreen2014}
{Elmegreen}, B.~G., {Struck}, C., \& {Hunter}, D.~A. 2014, \apj, 796, 110,
  \dodoi{10.1088/0004-637X/796/2/110}

\bibitem[{{Emami} {et~al.}(2021){Emami}, {Siana}, {El-Badry}, {Cook}, {Ma},
  {Weisz}, {Gharibshah}, {Alaee}, {Scarlata}, \& {Skillman}}]{Emami2021}
{Emami}, N., {Siana}, B., {El-Badry}, K., {et~al.} 2021, \apj, 922, 217,
  \dodoi{10.3847/1538-4357/ac1f8d}

\bibitem[{{Engler} {et~al.}(2023){Engler}, {Pillepich}, {Joshi}, {Pasquali},
  {Nelson}, \& {Grebel}}]{Engler2023}
{Engler}, C., {Pillepich}, A., {Joshi}, G.~D., {et~al.} 2023, \mnras, 522,
  5946, \dodoi{10.1093/mnras/stad1357}

\bibitem[{{Fall} \& {Efstathiou}(1980)}]{Fall1980}
{Fall}, S.~M., \& {Efstathiou}, G. 1980, \mnras, 193, 189,
  \dodoi{10.1093/mnras/193.2.189}

\bibitem[{{Fitts} {et~al.}(2018){Fitts}, {Boylan-Kolchin}, {Bullock}, {Weisz},
  {El-Badry}, {Wheeler}, {Faucher-Gigu{\`e}re}, {Quataert}, {Hopkins},
  {Kere{\v{s}}}, {Wetzel}, \& {Hayward}}]{Fitts2018}
{Fitts}, A., {Boylan-Kolchin}, M., {Bullock}, J.~S., {et~al.} 2018, \mnras,
  479, 319, \dodoi{10.1093/mnras/sty1488}

\bibitem[{{Frankel} {et~al.}(2018){Frankel}, {Rix}, {Ting}, {Ness}, \&
  {Hogg}}]{Frankel2018}
{Frankel}, N., {Rix}, H.-W., {Ting}, Y.-S., {Ness}, M., \& {Hogg}, D.~W. 2018,
  \apj, 865, 96, \dodoi{10.3847/1538-4357/aadba5}

\bibitem[{{Fry} {et~al.}(2015){Fry}, {Governato}, {Pontzen}, {Quinn},
  {Tremmel}, {Anderson}, {Menon}, {Brooks}, \& {Wadsley}}]{Fry2015}
{Fry}, A.~B., {Governato}, F., {Pontzen}, A., {et~al.} 2015, \mnras, 452, 1468,
  \dodoi{10.1093/mnras/stv1330}

\bibitem[{{Fu} {et~al.}(2024{\natexlab{a}}){Fu}, {Weisz}, {Starkenburg},
  {Martin}, {Mercado}, {Savino}, {Boylan-Kolchin}, {C{\^o}t{\'e}}, {Dolphin},
  {Longeard}, {Mateo}, {Samuel}, \& {Sandford}}]{Fu2024a}
{Fu}, S.~W., {Weisz}, D.~R., {Starkenburg}, E., {et~al.} 2024{\natexlab{a}},
  \apj, 965, 36, \dodoi{10.3847/1538-4357/ad25ed}

\bibitem[{{Fu} {et~al.}(2024{\natexlab{b}}){Fu}, {Weisz}, {Starkenburg},
  {Martin}, {Collins}, {Savino}, {Boylan-Kolchin}, {C{\^o}t{\'e}}, {Dolphin},
  {Longeard}, {Mateo}, {Mercado}, {Sandford}, \& {Skillman}}]{Fu2024b}
---. 2024{\natexlab{b}}, arXiv e-prints, arXiv:2407.04698,
  \dodoi{10.48550/arXiv.2407.04698}

\bibitem[{{Gallart} {et~al.}(2008){Gallart}, {Stetson}, {Meschin}, {Pont}, \&
  {Hardy}}]{Gallart2008}
{Gallart}, C., {Stetson}, P.~B., {Meschin}, I.~P., {Pont}, F., \& {Hardy}, E.
  2008, \apjl, 682, L89, \dodoi{10.1086/590552}

\bibitem[{{Genina} {et~al.}(2019){Genina}, {Frenk}, {Ben{\'\i}tez-Llambay},
  {Cole}, {Navarro}, {Oman}, \& {Fattahi}}]{Genina2019}
{Genina}, A., {Frenk}, C.~S., {Ben{\'\i}tez-Llambay}, A., {et~al.} 2019,
  \mnras, 488, 2312, \dodoi{10.1093/mnras/stz1852}

\bibitem[{{Gill} {et~al.}(2004){Gill}, {Knebe}, \& {Gibson}}]{Gill_2004}
{Gill}, S. P.~D., {Knebe}, A., \& {Gibson}, B.~K. 2004, \mnras, 351, 399,
  \dodoi{10.1111/j.1365-2966.2004.07786.x}

\bibitem[{{Girardi} {et~al.}(2010){Girardi}, {Williams}, {Gilbert},
  {Rosenfield}, {Dalcanton}, {Marigo}, {Boyer}, {Dolphin}, {Weisz},
  {Melbourne}, {Olsen}, {Seth}, \& {Skillman}}]{Girardi2010}
{Girardi}, L., {Williams}, B.~F., {Gilbert}, K.~M., {et~al.} 2010, \apj, 724,
  1030, \dodoi{10.1088/0004-637X/724/2/1030}

\bibitem[{{Gnedin}(2012)}]{Gnedin2012}
{Gnedin}, N.~Y. 2012, \apj, 754, 113, \dodoi{10.1088/0004-637X/754/2/113}

\bibitem[{{Goddard} {et~al.}(2017){Goddard}, {Thomas}, {Maraston}, {Westfall},
  {Etherington}, {Riffel}, {Mallmann}, {Zheng}, {Argudo-Fern{\'a}ndez},
  {Bershady}, {Bundy}, {Drory}, {Law}, {Yan}, {Wake}, {Weijmans}, {Bizyaev},
  {Brownstein}, {Lane}, {Maiolino}, {Masters}, {Merrifield}, {Nitschelm},
  {Pan}, {Roman-Lopes}, \& {Storchi-Bergmann}}]{Goddard2017}
{Goddard}, D., {Thomas}, D., {Maraston}, C., {et~al.} 2017, \mnras, 465, 688,
  \dodoi{10.1093/mnras/stw2719}

\bibitem[{{Gogarten} {et~al.}(2010){Gogarten}, {Dalcanton}, {Williams},
  {Ro{\v{s}}kar}, {Holtzman}, {Seth}, {Dolphin}, {Weisz}, {Cole}, {Debattista},
  {Gilbert}, {Olsen}, {Skillman}, {de Jong}, {Karachentsev}, \&
  {Quinn}}]{Gogarten2010}
{Gogarten}, S.~M., {Dalcanton}, J.~J., {Williams}, B.~F., {et~al.} 2010, \apj,
  712, 858, \dodoi{10.1088/0004-637X/712/2/858}

\bibitem[{{Gonz{\'a}lez Delgado} {et~al.}(2014){Gonz{\'a}lez Delgado},
  {P{\'e}rez}, {Cid Fernandes}, {Garc{\'\i}a-Benito}, {de Amorim},
  {S{\'a}nchez}, {Husemann}, {Cortijo-Ferrero}, {L{\'o}pez Fern{\'a}ndez},
  {S{\'a}nchez-Bl{\'a}zquez}, {Bekeraite}, {Walcher}, {Falc{\'o}n-Barroso},
  {Gallazzi}, {van de Ven}, {Alves}, {Bland-Hawthorn}, {Kennicutt}, {Kupko},
  {Lyubenova}, {Mast}, {Moll{\'a}}, {Marino}, {Quirrenbach}, {V{\'\i}lchez}, \&
  {Wisotzki}}]{Delgado2014}
{Gonz{\'a}lez Delgado}, R.~M., {P{\'e}rez}, E., {Cid Fernandes}, R., {et~al.}
  2014, \aap, 562, A47, \dodoi{10.1051/0004-6361/201322011}

\bibitem[{{Gonz{\'a}lez Delgado} {et~al.}(2015){Gonz{\'a}lez Delgado},
  {Garc{\'\i}a-Benito}, {P{\'e}rez}, {Cid Fernandes}, {de Amorim},
  {Cortijo-Ferrero}, {Lacerda}, {L{\'o}pez Fern{\'a}ndez}, {Vale-Asari},
  {S{\'a}nchez}, {Moll{\'a}}, {Ruiz-Lara}, {S{\'a}nchez-Bl{\'a}zquez},
  {Walcher}, {Alves}, {Aguerri}, {Bekerait{\'e}}, {Bland-Hawthorn}, {Galbany},
  {Gallazzi}, {Husemann}, {Iglesias-P{\'a}ramo}, {Kalinova},
  {L{\'o}pez-S{\'a}nchez}, {Marino}, {M{\'a}rquez}, {Masegosa}, {Mast},
  {M{\'e}ndez-Abreu}, {Mendoza}, {del Olmo}, {P{\'e}rez}, {Quirrenbach}, \&
  {Zibetti}}]{Delgado2015}
{Gonz{\'a}lez Delgado}, R.~M., {Garc{\'\i}a-Benito}, R., {P{\'e}rez}, E.,
  {et~al.} 2015, \aap, 581, A103, \dodoi{10.1051/0004-6361/201525938}

\bibitem[{{Governato} {et~al.}(2010){Governato}, {Brook}, {Mayer}, {Brooks},
  {Rhee}, {Wadsley}, {Jonsson}, {Willman}, {Stinson}, {Quinn}, \&
  {Madau}}]{Governato2010}
{Governato}, F., {Brook}, C., {Mayer}, L., {et~al.} 2010, \nat, 463, 203,
  \dodoi{10.1038/nature08640}

\bibitem[{{Governato} {et~al.}(2012){Governato}, {Zolotov}, {Pontzen},
  {Christensen}, {Oh}, {Brooks}, {Quinn}, {Shen}, \& {Wadsley}}]{Governato2012}
{Governato}, F., {Zolotov}, A., {Pontzen}, A., {et~al.} 2012, \mnras, 422,
  1231, \dodoi{10.1111/j.1365-2966.2012.20696.x}

\bibitem[{Graus {et~al.}(2019)Graus, Bullock, Fitts, Cooper, Boylan-Kolchin,
  Weisz, Wetzel, Feldmann, Faucher-Giguère, Quataert, \& et~al.}]{Graus_2019}
Graus, A.~S., Bullock, J.~S., Fitts, A., {et~al.} 2019, Monthly Notices of the
  Royal Astronomical Society, 490, 1186–1201, \dodoi{10.1093/mnras/stz2649}

\bibitem[{{Haardt} \& {Madau}(2012)}]{Haardt2012}
{Haardt}, F., \& {Madau}, P. 2012, \apj, 746, 125,
  \dodoi{10.1088/0004-637X/746/2/125}

\bibitem[{{Han} {et~al.}(2020){Han}, {Kim}, {Yoon}, {Lee}, {Arimoto},
  {Okamoto}, \& {Ree}}]{Han2020}
{Han}, S.-I., {Kim}, H.-S., {Yoon}, S.-J., {et~al.} 2020, \apjs, 247, 7,
  \dodoi{10.3847/1538-4365/ab6441}

\bibitem[{{Harbeck} {et~al.}(2001){Harbeck}, {Grebel}, {Holtzman},
  {Guhathakurta}, {Brandner}, {Geisler}, {Sarajedini}, {Dolphin},
  {Hurley-Keller}, \& {Mateo}}]{Harbeck2001}
{Harbeck}, D., {Grebel}, E.~K., {Holtzman}, J., {et~al.} 2001, \aj, 122, 3092,
  \dodoi{10.1086/324232}

\bibitem[{{Hidalgo} {et~al.}(2009){Hidalgo}, {Aparicio}, \&
  {Gallart}}]{Hidalgo2009}
{Hidalgo}, S.~L., {Aparicio}, A., \& {Gallart}, C. 2009, in The Ages of Stars,
  ed. E.~E. {Mamajek}, D.~R. {Soderblom}, \& R.~F.~G. {Wyse}, Vol. 258,
  245--252, \dodoi{10.1017/S1743921309031895}

\bibitem[{{Hidalgo} {et~al.}(2013){Hidalgo}, {Monelli}, {Aparicio}, {Gallart},
  {Skillman}, {Cassisi}, {Bernard}, {Mayer}, {Stetson}, {Cole}, \&
  {Dolphin}}]{Hidalgo2013}
{Hidalgo}, S.~L., {Monelli}, M., {Aparicio}, A., {et~al.} 2013, \apj, 778, 103,
  \dodoi{10.1088/0004-637X/778/2/103}

\bibitem[{{Hirtenstein} {et~al.}(2019){Hirtenstein}, {Jones}, {Wang}, {Wetzel},
  {El-Badry}, {Hoag}, {Treu}, {Brada{\v{c}}}, \& {Morishita}}]{Hirtenstein2019}
{Hirtenstein}, J., {Jones}, T., {Wang}, X., {et~al.} 2019, \apj, 880, 54,
  \dodoi{10.3847/1538-4357/ab113e}

\bibitem[{{Hohl}(1971)}]{Hohl1971}
{Hohl}, F. 1971, \apj, 168, 343, \dodoi{10.1086/151091}

\bibitem[{Hunter \& Elmegreen(2006)}]{Hunter_2006}
Hunter, D.~A., \& Elmegreen, B.~G. 2006, The Astrophysical Journal Supplement
  Series, 162, 49, \dodoi{10.1086/498096}

\bibitem[{{Indu} \& {Subramaniam}(2011)}]{Indu2011}
{Indu}, G., \& {Subramaniam}, A. 2011, \aap, 535, A115,
  \dodoi{10.1051/0004-6361/201117298}

\bibitem[{{Jackson} {et~al.}(2024){Jackson}, {Kaviraj}, {Yi}, {Peirani},
  {Dubois}, {Martin}, {Devriendt}, {Slyz}, {Pichon}, {Volonteri}, {Kimm}, \&
  {Kraljic}}]{Jackson2024}
{Jackson}, R.~A., {Kaviraj}, S., {Yi}, S.~K., {et~al.} 2024, \mnras, 528, 1655,
  \dodoi{10.1093/mnras/stae056}

\bibitem[{Jansen {et~al.}(2000)Jansen, Fabricant, Franx, \&
  Caldwell}]{Jansen_2000}
Jansen, R.~A., Fabricant, D., Franx, M., \& Caldwell, N. 2000, The
  Astrophysical Journal Supplement Series, 126, 331–397,
  \dodoi{10.1086/313308}

\bibitem[{{Javadi} {et~al.}(2017){Javadi}, {van Loon}, {Khosroshahi},
  {Tabatabaei}, {Hamedani Golshan}, \& {Rashidi}}]{Javadi2017}
{Javadi}, A., {van Loon}, J.~T., {Khosroshahi}, H.~G., {et~al.} 2017, \mnras,
  464, 2103, \dodoi{10.1093/mnras/stw2463}

\bibitem[{{Joshi} {et~al.}(2021){Joshi}, {Pillepich}, {Nelson}, {Zinger},
  {Marinacci}, {Springel}, {Vogelsberger}, \& {Hernquist}}]{Joshi2021}
{Joshi}, G.~D., {Pillepich}, A., {Nelson}, D., {et~al.} 2021, \mnras, 508,
  1652, \dodoi{10.1093/mnras/stab2573}

\bibitem[{{Katz} \& {White}(1993)}]{Katz_1993}
{Katz}, N., \& {White}, S. D.~M. 1993, \apj, 412, 455, \dodoi{10.1086/172935}

\bibitem[{{Kawata} {et~al.}(2006){Kawata}, {Arimoto}, {Cen}, \&
  {Gibson}}]{Kawata2006}
{Kawata}, D., {Arimoto}, N., {Cen}, R., \& {Gibson}, B.~K. 2006, \apj, 641,
  785, \dodoi{10.1086/500633}

\bibitem[{Kepner(1999)}]{Kepner_1999}
Kepner, J.~V. 1999, The Astrophysical Journal, 520, 59–66,
  \dodoi{10.1086/307419}

\bibitem[{{Knollmann} \& {Knebe}(2009)}]{Knollmann_2009}
{Knollmann}, S.~R., \& {Knebe}, A. 2009, \apjs, 182, 608,
  \dodoi{10.1088/0067-0049/182/2/608}

\bibitem[{{Koleva} {et~al.}(2011){Koleva}, {Prugniel}, {De Rijcke}, \&
  {Zeilinger}}]{Koleva2011}
{Koleva}, M., {Prugniel}, P., {De Rijcke}, S., \& {Zeilinger}, W.~W. 2011,
  \mnras, 417, 1643, \dodoi{10.1111/j.1365-2966.2011.19057.x}

\bibitem[{{Kroupa}(2001)}]{Kroupa2001}
{Kroupa}, P. 2001, \mnras, 322, 231, \dodoi{10.1046/j.1365-8711.2001.04022.x}

\bibitem[{{Larson}(1976)}]{Larson_1976}
{Larson}, R.~B. 1976, \mnras, 176, 31, \dodoi{10.1093/mnras/176.1.31}

\bibitem[{{Lazar} {et~al.}(2020){Lazar}, {Bullock}, {Boylan-Kolchin}, {Chan},
  {Hopkins}, {Graus}, {Wetzel}, {El-Badry}, {Wheeler}, {Straight},
  {Kere{\v{s}}}, {Faucher-Gigu{\`e}re}, {Fitts}, \&
  {Garrison-Kimmel}}]{Lazar_2020}
{Lazar}, A., {Bullock}, J.~S., {Boylan-Kolchin}, M., {et~al.} 2020, \mnras,
  497, 2393, \dodoi{10.1093/mnras/staa2101}

\bibitem[{{Liao} \& {Cooper}(2023)}]{Liao2023}
{Liao}, L.-W., \& {Cooper}, A.~P. 2023, \mnras, 518, 3999,
  \dodoi{10.1093/mnras/stac3327}

\bibitem[{{Ludlow} {et~al.}(2020){Ludlow}, {Schaye}, {Schaller}, \&
  {Bower}}]{Ludlow2020}
{Ludlow}, A.~D., {Schaye}, J., {Schaller}, M., \& {Bower}, R. 2020, \mnras,
  493, 2926, \dodoi{10.1093/mnras/staa316}

\bibitem[{{Ludlow} {et~al.}(2019){Ludlow}, {Schaye}, {Schaller}, \&
  {Richings}}]{Ludlow2019}
{Ludlow}, A.~D., {Schaye}, J., {Schaller}, M., \& {Richings}, J. 2019, \mnras,
  488, L123, \dodoi{10.1093/mnrasl/slz110}

\bibitem[{{Ma} {et~al.}(2024){Ma}, {Du}, {Ho}, {Sheng}, \& {Liao}}]{Ma2024}
{Ma}, H.-C., {Du}, M., {Ho}, L.~C., {Sheng}, M.-j., \& {Liao}, S. 2024, arXiv
  e-prints, arXiv:2404.10432, \dodoi{10.48550/arXiv.2404.10432}

\bibitem[{{Marigo} {et~al.}(2008){Marigo}, {Girardi}, {Bressan}, {Groenewegen},
  {Silva}, \& {Granato}}]{Marigo2008}
{Marigo}, P., {Girardi}, L., {Bressan}, A., {et~al.} 2008, \aap, 482, 883,
  \dodoi{10.1051/0004-6361:20078467}

\bibitem[{{Mart{\'\i}nez-V{\'a}zquez}
  {et~al.}(2015){Mart{\'\i}nez-V{\'a}zquez}, {Monelli}, {Bono}, {Stetson},
  {Ferraro}, {Bernard}, {Gallart}, {Fiorentino}, {Iannicola}, \&
  {Udalski}}]{Martinez2015}
{Mart{\'\i}nez-V{\'a}zquez}, C.~E., {Monelli}, M., {Bono}, G., {et~al.} 2015,
  \mnras, 454, 1509, \dodoi{10.1093/mnras/stv2014}

\bibitem[{{Mart{\'\i}nez-V{\'a}zquez}
  {et~al.}(2021){Mart{\'\i}nez-V{\'a}zquez}, {Monelli}, {Cassisi}, {Taibi},
  {Gallart}, {Vivas}, {Walker}, {Mart{\'\i}n-Ravelo}, {Zenteno}, {Battaglia},
  {Bono}, {Calamida}, {Carollo}, {Cicu{\'e}ndez}, {Fiorentino}, {Marconi},
  {Salvadori}, {Balbinot}, {Bernard}, {Dall'Ora}, \& {Stetson}}]{Martinez2021}
{Mart{\'\i}nez-V{\'a}zquez}, C.~E., {Monelli}, M., {Cassisi}, S., {et~al.}
  2021, \mnras, 508, 1064, \dodoi{10.1093/mnras/stab2493}

\bibitem[{{Maxwell} {et~al.}(2012){Maxwell}, {Wadsley}, {Couchman}, \&
  {Mashchenko}}]{Maxwell2012}
{Maxwell}, A.~J., {Wadsley}, J., {Couchman}, H.~M.~P., \& {Mashchenko}, S.
  2012, \apjl, 755, L35, \dodoi{10.1088/2041-8205/755/2/L35}

\bibitem[{Mayer {et~al.}(2006)Mayer, Mastropietro, Wadsley, Stadel, \&
  Moore}]{Mayer_2006}
Mayer, L., Mastropietro, C., Wadsley, J., Stadel, J., \& Moore, B. 2006,
  Monthly Notices of the Royal Astronomical Society, 369, 1021–1038,
  \dodoi{10.1111/j.1365-2966.2006.10403.x}

\bibitem[{{McConnachie} {et~al.}(2006){McConnachie}, {Arimoto}, {Irwin}, \&
  {Tolstoy}}]{McConnachie2006}
{McConnachie}, A.~W., {Arimoto}, N., {Irwin}, M., \& {Tolstoy}, E. 2006,
  \mnras, 373, 715, \dodoi{10.1111/j.1365-2966.2006.11053.x}

\bibitem[{{McKee} \& {Ostriker}(1977)}]{McKee1977}
{McKee}, C.~F., \& {Ostriker}, J.~P. 1977, \apj, 218, 148,
  \dodoi{10.1086/155667}

\bibitem[{{McQuinn} {et~al.}(2010){McQuinn}, {Skillman}, {Cannon}, {Dalcanton},
  {Dolphin}, {Hidalgo-Rodr{\'\i}guez}, {Holtzman}, {Stark}, {Weisz}, \&
  {Williams}}]{McQuinn2010}
{McQuinn}, K. B.~W., {Skillman}, E.~D., {Cannon}, J.~M., {et~al.} 2010, \apj,
  724, 49, \dodoi{10.1088/0004-637X/724/1/49}

\bibitem[{{McQuinn} {et~al.}(2017){McQuinn}, {Boyer}, {Mitchell}, {Skillman},
  {Gehrz}, {Groenewegen}, {McDonald}, {Sloan}, {van Loon}, {Whitelock}, \&
  {Zijlstra}}]{McQuinn2017}
{McQuinn}, K. B.~W., {Boyer}, M.~L., {Mitchell}, M.~B., {et~al.} 2017, \apj,
  834, 78, \dodoi{10.3847/1538-4357/834/1/78}

\bibitem[{{Menon} {et~al.}(2015){Menon}, {Wesolowski}, {Zheng}, {Jetley},
  {Kale}, {Quinn}, \& {Governato}}]{Menon_2015}
{Menon}, H., {Wesolowski}, L., {Zheng}, G., {et~al.} 2015, Computational
  Astrophysics and Cosmology, 2, 1, \dodoi{10.1186/s40668-015-0007-9}

\bibitem[{{Mercado} {et~al.}(2021){Mercado}, {Bullock}, {Boylan-Kolchin},
  {Moreno}, {Wetzel}, {El-Badry}, {Graus}, {Fitts}, {Hopkins},
  {Faucher-Gigu{\`e}re}, \& {Gurvich}}]{Mercado2021}
{Mercado}, F.~J., {Bullock}, J.~S., {Boylan-Kolchin}, M., {et~al.} 2021,
  \mnras, 501, 5121, \dodoi{10.1093/mnras/staa3958}

\bibitem[{{Morelli} {et~al.}(2015){Morelli}, {Corsini}, {Pizzella}, {Dalla
  Bont{\`a}}, {Coccato}, \& {M{\'e}ndez-Abreu}}]{Morelli2015}
{Morelli}, L., {Corsini}, E.~M., {Pizzella}, A., {et~al.} 2015, \mnras, 452,
  1128, \dodoi{10.1093/mnras/stv1357}

\bibitem[{{Mostoghiu} {et~al.}(2018){Mostoghiu}, {Di Cintio}, {Knebe},
  {Libeskind}, {Minchev}, \& {Brook}}]{Mostoghiu2018}
{Mostoghiu}, R., {Di Cintio}, A., {Knebe}, A., {et~al.} 2018, \mnras, 480,
  4455, \dodoi{10.1093/mnras/sty2161}

\bibitem[{{Muni} {et~al.}(2024){Muni}, {Pontzen}, {Read}, {Agertz}, {Rey}, \&
  {Taylor}}]{Muni2024}
{Muni}, C., {Pontzen}, A., {Read}, J.~I., {et~al.} 2024, arXiv e-prints,
  arXiv:2407.14579, \dodoi{10.48550/arXiv.2407.14579}

\bibitem[{{Munshi} {et~al.}(2021){Munshi}, {Brooks}, {Applebaum},
  {Christensen}, {Quinn}, \& {Sligh}}]{Munshi_2021}
{Munshi}, F., {Brooks}, A.~M., {Applebaum}, E., {et~al.} 2021, \apj, 923, 35,
  \dodoi{10.3847/1538-4357/ac0db6}

\bibitem[{{Munshi} {et~al.}(2013){Munshi}, {Governato}, {Brooks},
  {Christensen}, {Shen}, {Loebman}, {Moster}, {Quinn}, \&
  {Wadsley}}]{Munshi2013}
{Munshi}, F., {Governato}, F., {Brooks}, A.~M., {et~al.} 2013, \apj, 766, 56,
  \dodoi{10.1088/0004-637X/766/1/56}

\bibitem[{{Navarro} {et~al.}(1996){Navarro}, {Eke}, \& {Frenk}}]{Navarro1996}
{Navarro}, J.~F., {Eke}, V.~R., \& {Frenk}, C.~S. 1996, \mnras, 283, L72,
  \dodoi{10.1093/mnras/283.3.L72}

\bibitem[{{O{\~n}orbe} {et~al.}(2014){O{\~n}orbe}, {Garrison-Kimmel}, {Maller},
  {Bullock}, {Rocha}, \& {Hahn}}]{Onorbe2014}
{O{\~n}orbe}, J., {Garrison-Kimmel}, S., {Maller}, A.~H., {et~al.} 2014,
  \mnras, 437, 1894, \dodoi{10.1093/mnras/stt2020}

\bibitem[{{Oh} {et~al.}(2011){Oh}, {Brook}, {Governato}, {Brinks}, {Mayer}, {de
  Blok}, {Brooks}, \& {Walter}}]{Oh2011}
{Oh}, S.-H., {Brook}, C., {Governato}, F., {et~al.} 2011, \aj, 142, 24,
  \dodoi{10.1088/0004-6256/142/1/24}

\bibitem[{{Okamoto} {et~al.}(2017){Okamoto}, {Arimoto}, {Tolstoy}, {Jablonka},
  {Irwin}, {Komiyama}, {Yamada}, \& {Onodera}}]{Okamoto2017}
{Okamoto}, S., {Arimoto}, N., {Tolstoy}, E., {et~al.} 2017, \mnras, 467, 208,
  \dodoi{10.1093/mnras/stx086}

\bibitem[{{Patel} {et~al.}(2023){Patel}, {Kelson}, {Abramson}, {Sattari}, \&
  {Lorenz}}]{Patel2023}
{Patel}, S.~G., {Kelson}, D.~D., {Abramson}, L.~E., {Sattari}, Z., \& {Lorenz},
  B. 2023, \apj, 945, 93, \dodoi{10.3847/1538-4357/acb938}

\bibitem[{{Patel} {et~al.}(2018){Patel}, {Kelson}, {Diao}, {Tonnesen}, \&
  {Abramson}}]{Patel2018}
{Patel}, S.~G., {Kelson}, D.~D., {Diao}, N., {Tonnesen}, S., \& {Abramson},
  L.~E. 2018, \apjl, 866, L21, \dodoi{10.3847/2041-8213/aae524}

\bibitem[{{Pelliccia} {et~al.}(2020){Pelliccia}, {Mobasher}, {Darvish},
  {Lemaux}, {Lubin}, {Hirtenstein}, {Shen}, {Wu}, {El-Badry}, {Wetzel}, \&
  {Jones}}]{Pelliccia2020}
{Pelliccia}, D., {Mobasher}, B., {Darvish}, B., {et~al.} 2020, \apjl, 896, L26,
  \dodoi{10.3847/2041-8213/ab9815}

\bibitem[{{Pessa} {et~al.}(2023){Pessa}, {Schinnerer}, {Sanchez-Blazquez},
  {Belfiore}, {Groves}, {Emsellem}, {Neumann}, {Leroy}, {Bigiel}, {Chevance},
  {Dale}, {Glover}, {Grasha}, {Klessen}, {Kreckel}, {Kruijssen}, {Pinna},
  {Querejeta}, {Rosolowsky}, \& {Williams}}]{Pessa2023}
{Pessa}, I., {Schinnerer}, E., {Sanchez-Blazquez}, P., {et~al.} 2023, \aap,
  673, A147, \dodoi{10.1051/0004-6361/202245673}

\bibitem[{{Peterken} {et~al.}(2020){Peterken}, {Merrifield},
  {Arag{\'o}n-Salamanca}, {Fraser-McKelvie}, {Avila-Reese}, {Riffel}, {Knapen},
  \& {Drory}}]{Peterken2020}
{Peterken}, T., {Merrifield}, M., {Arag{\'o}n-Salamanca}, A., {et~al.} 2020,
  \mnras, 495, 3387, \dodoi{10.1093/mnras/staa1303}

\bibitem[{{Piatti}(2018)}]{Piatti2018}
{Piatti}, A.~E. 2018, \mnras, 473, 4410, \dodoi{10.1093/mnras/stx2686}

\bibitem[{{Planck Collaboration} {et~al.}(2016){Planck Collaboration}, {Ade},
  {Aghanim}, {Arnaud}, {Ashdown}, {Aumont}, {Baccigalupi}, {Banday},
  {Barreiro}, {Bartlett}, {Bartolo}, {Battaner}, {Battye}, {Benabed},
  {Beno{\^\i}t}, {Benoit-L{\'e}vy}, {Bernard}, {Bersanelli}, {Bielewicz},
  {Bock}, {Bonaldi}, {Bonavera}, {Bond}, {Borrill}, {Bouchet}, {Boulanger},
  {Bucher}, {Burigana}, {Butler}, {Calabrese}, {Cardoso}, {Catalano},
  {Challinor}, {Chamballu}, {Chary}, {Chiang}, {Chluba}, {Christensen},
  {Church}, {Clements}, {Colombi}, {Colombo}, {Combet}, {Coulais}, {Crill},
  {Curto}, {Cuttaia}, {Danese}, {Davies}, {Davis}, {de Bernardis}, {de Rosa},
  {de Zotti}, {Delabrouille}, {D{\'e}sert}, {Di Valentino}, {Dickinson},
  {Diego}, {Dolag}, {Dole}, {Donzelli}, {Dor{\'e}}, {Douspis}, {Ducout},
  {Dunkley}, {Dupac}, {Efstathiou}, {Elsner}, {En{\ss}lin}, {Eriksen},
  {Farhang}, {Fergusson}, {Finelli}, {Forni}, {Frailis}, {Fraisse},
  {Franceschi}, {Frejsel}, {Galeotta}, {Galli}, {Ganga}, {Gauthier}, {Gerbino},
  {Ghosh}, {Giard}, {Giraud-H{\'e}raud}, {Giusarma}, {Gjerl{\o}w},
  {Gonz{\'a}lez-Nuevo}, {G{\'o}rski}, {Gratton}, {Gregorio}, {Gruppuso},
  {Gudmundsson}, {Hamann}, {Hansen}, {Hanson}, {Harrison}, {Helou},
  {Henrot-Versill{\'e}}, {Hern{\'a}ndez-Monteagudo}, {Herranz}, {Hildebrandt},
  {Hivon}, {Hobson}, {Holmes}, {Hornstrup}, {Hovest}, {Huang}, {Huffenberger},
  {Hurier}, {Jaffe}, {Jaffe}, {Jones}, {Juvela}, {Keih{\"a}nen}, {Keskitalo},
  {Kisner}, {Kneissl}, {Knoche}, {Knox}, {Kunz}, {Kurki-Suonio}, {Lagache},
  {L{\"a}hteenm{\"a}ki}, {Lamarre}, {Lasenby}, {Lattanzi}, {Lawrence}, {Leahy},
  {Leonardi}, {Lesgourgues}, {Levrier}, {Lewis}, {Liguori}, {Lilje},
  {Linden-V{\o}rnle}, {L{\'o}pez-Caniego}, {Lubin}, {Mac{\'\i}as-P{\'e}rez},
  {Maggio}, {Maino}, {Mandolesi}, {Mangilli}, {Marchini}, {Maris}, {Martin},
  {Martinelli}, {Mart{\'\i}nez-Gonz{\'a}lez}, {Masi}, {Matarrese}, {McGehee},
  {Meinhold}, {Melchiorri}, {Melin}, {Mendes}, {Mennella}, {Migliaccio},
  {Millea}, {Mitra}, {Miville-Desch{\^e}nes}, {Moneti}, {Montier}, {Morgante},
  {Mortlock}, {Moss}, {Munshi}, {Murphy}, {Naselsky}, {Nati}, {Natoli},
  {Netterfield}, {N{\o}rgaard-Nielsen}, {Noviello}, {Novikov}, {Novikov},
  {Oxborrow}, {Paci}, {Pagano}, {Pajot}, {Paladini}, {Paoletti}, {Partridge},
  {Pasian}, {Patanchon}, {Pearson}, {Perdereau}, {Perotto}, {Perrotta},
  {Pettorino}, {Piacentini}, {Piat}, {Pierpaoli}, {Pietrobon}, {Plaszczynski},
  {Pointecouteau}, {Polenta}, {Popa}, {Pratt}, {Pr{\'e}zeau}, {Prunet},
  {Puget}, {Rachen}, {Reach}, {Rebolo}, {Reinecke}, {Remazeilles}, {Renault},
  {Renzi}, {Ristorcelli}, {Rocha}, {Rosset}, {Rossetti}, {Roudier},
  {Rouill{\'e} d'Orfeuil}, {Rowan-Robinson}, {Rubi{\~n}o-Mart{\'\i}n},
  {Rusholme}, {Said}, {Salvatelli}, {Salvati}, {Sandri}, {Santos},
  {Savelainen}, {Savini}, {Scott}, {Seiffert}, {Serra}, {Shellard}, {Spencer},
  {Spinelli}, {Stolyarov}, {Stompor}, {Sudiwala}, {Sunyaev}, {Sutton},
  {Suur-Uski}, {Sygnet}, {Tauber}, {Terenzi}, {Toffolatti}, {Tomasi},
  {Tristram}, {Trombetti}, {Tucci}, {Tuovinen}, {T{\"u}rler}, {Umana},
  {Valenziano}, {Valiviita}, {Van Tent}, {Vielva}, {Villa}, {Wade}, {Wandelt},
  {Wehus}, {White}, {White}, {Wilkinson}, {Yvon}, {Zacchei}, \&
  {Zonca}}]{Planck2016}
{Planck Collaboration}, {Ade}, P.~A.~R., {Aghanim}, N., {et~al.} 2016, \aap,
  594, A13, \dodoi{10.1051/0004-6361/201525830}

\bibitem[{{Pontzen} \& {Governato}(2012)}]{Pontzen2012}
{Pontzen}, A., \& {Governato}, F. 2012, \mnras, 421, 3464,
  \dodoi{10.1111/j.1365-2966.2012.20571.x}

\bibitem[{{Power} {et~al.}(2003){Power}, {Navarro}, {Jenkins}, {Frenk},
  {White}, {Springel}, {Stadel}, \& {Quinn}}]{Power_2003}
{Power}, C., {Navarro}, J.~F., {Jenkins}, A., {et~al.} 2003, \mnras, 338, 14,
  \dodoi{10.1046/j.1365-8711.2003.05925.x}

\bibitem[{{Prescott} {et~al.}(2011){Prescott}, {Baldry}, {James}, {Bamford},
  {Bland-Hawthorn}, {Brough}, {Brown}, {Cameron}, {Conselice}, {Croom},
  {Driver}, {Frenk}, {Gunawardhana}, {Hill}, {Hopkins}, {Jones}, {Kelvin},
  {Kuijken}, {Liske}, {Loveday}, {Nichol}, {Norberg}, {Parkinson}, {Peacock},
  {Phillipps}, {Pimbblet}, {Popescu}, {Robotham}, {Sharp}, {Sutherland},
  {Taylor}, {Tuffs}, {van Kampen}, \& {Wijesinghe}}]{Prescott2011}
{Prescott}, M., {Baldry}, I.~K., {James}, P.~A., {et~al.} 2011, \mnras, 417,
  1374, \dodoi{10.1111/j.1365-2966.2011.19353.x}

\bibitem[{{Radburn-Smith} {et~al.}(2012){Radburn-Smith}, {Ro{\v{s}}kar},
  {Debattista}, {Dalcanton}, {Streich}, {de Jong}, {Vlaji{\'c}}, {Holwerda},
  {Purcell}, {Dolphin}, \& {Zucker}}]{Radburn-Smith2012}
{Radburn-Smith}, D.~J., {Ro{\v{s}}kar}, R., {Debattista}, V.~P., {et~al.} 2012,
  \apj, 753, 138, \dodoi{10.1088/0004-637X/753/2/138}

\bibitem[{{Read} \& {Gilmore}(2005)}]{Read2005}
{Read}, J.~I., \& {Gilmore}, G. 2005, \mnras, 356, 107,
  \dodoi{10.1111/j.1365-2966.2004.08424.x}

\bibitem[{{Read} {et~al.}(2019){Read}, {Walker}, \& {Steger}}]{Read2019}
{Read}, J.~I., {Walker}, M.~G., \& {Steger}, P. 2019, \mnras, 484, 1401,
  \dodoi{10.1093/mnras/sty3404}

\bibitem[{{Relatores} {et~al.}(2019){Relatores}, {Newman}, {Simon}, {Ellis},
  {Truong}, {Blitz}, {Bolatto}, {Martin}, {Matuszewski}, {Morrissey}, \&
  {Neill}}]{Relatores2019}
{Relatores}, N.~C., {Newman}, A.~B., {Simon}, J.~D., {et~al.} 2019, \apj, 887,
  94, \dodoi{10.3847/1538-4357/ab5305}

\bibitem[{{Ro{\v{s}}kar} {et~al.}(2008){Ro{\v{s}}kar}, {Debattista}, {Quinn},
  {Stinson}, \& {Wadsley}}]{Roskar2008}
{Ro{\v{s}}kar}, R., {Debattista}, V.~P., {Quinn}, T.~R., {Stinson}, G.~S., \&
  {Wadsley}, J. 2008, \apjl, 684, L79, \dodoi{10.1086/592231}

\bibitem[{{Sales} {et~al.}(2012){Sales}, {Navarro}, {Theuns}, {Schaye},
  {White}, {Frenk}, {Crain}, \& {Dalla Vecchia}}]{Sales2012}
{Sales}, L.~V., {Navarro}, J.~F., {Theuns}, T., {et~al.} 2012, \mnras, 423,
  1544, \dodoi{10.1111/j.1365-2966.2012.20975.x}

\bibitem[{{Sellwood} \& {Binney}(2002)}]{Sellwood2002}
{Sellwood}, J.~A., \& {Binney}, J.~J. 2002, \mnras, 336, 785,
  \dodoi{10.1046/j.1365-8711.2002.05806.x}

\bibitem[{{Sharina} {et~al.}(2018){Sharina}, {Makarova}, \&
  {Makarov}}]{Sharina2023}
{Sharina}, M.~E., {Makarova}, L.~N., \& {Makarov}, D.~I. 2018, Astrophysics,
  61, 435, \dodoi{10.1007/s10511-018-9548-3}

\bibitem[{{Shen} {et~al.}(2010){Shen}, {Wadsley}, \& {Stinson}}]{Shen2010}
{Shen}, S., {Wadsley}, J., \& {Stinson}, G. 2010, \mnras, 407, 1581,
  \dodoi{10.1111/j.1365-2966.2010.17047.x}

\bibitem[{{Simpson} {et~al.}(2018){Simpson}, {Grand}, {G{\'o}mez}, {Marinacci},
  {Pakmor}, {Springel}, {Campbell}, \& {Frenk}}]{Simpson2018}
{Simpson}, C.~M., {Grand}, R. J.~J., {G{\'o}mez}, F.~A., {et~al.} 2018, \mnras,
  478, 548, \dodoi{10.1093/mnras/sty774}

\bibitem[{{Spergel} {et~al.}(2007){Spergel}, {Bean}, {Dor{\'e}}, {Nolta},
  {Bennett}, {Dunkley}, {Hinshaw}, {Jarosik}, {Komatsu}, {Page}, {Peiris},
  {Verde}, {Halpern}, {Hill}, {Kogut}, {Limon}, {Meyer}, {Odegard}, {Tucker},
  {Weiland}, {Wollack}, \& {Wright}}]{Spergel_2007}
{Spergel}, D.~N., {Bean}, R., {Dor{\'e}}, O., {et~al.} 2007, \apjs, 170, 377,
  \dodoi{10.1086/513700}

\bibitem[{{Stadel}(2001)}]{Stadel2001}
{Stadel}, J.~G. 2001, PhD thesis, University of Washington, Seattle

\bibitem[{{Stinson} {et~al.}(2006){Stinson}, {Seth}, {Katz}, {Wadsley},
  {Governato}, \& {Quinn}}]{Stinson_2006}
{Stinson}, G., {Seth}, A., {Katz}, N., {et~al.} 2006, \mnras, 373, 1074,
  \dodoi{10.1111/j.1365-2966.2006.11097.x}

\bibitem[{Stinson {et~al.}(2009)Stinson, Dalcanton, Quinn, Gogarten, Kaufmann,
  \& Wadsley}]{Stinson_2009}
Stinson, G.~S., Dalcanton, J.~J., Quinn, T., {et~al.} 2009, Monthly Notices of
  the Royal Astronomical Society, 395, 1455–1466,
  \dodoi{10.1111/j.1365-2966.2009.14555.x}

\bibitem[{{Taibi} {et~al.}(2022){Taibi}, {Battaglia}, {Leaman}, {Brooks},
  {Riggs}, {Munshi}, {Revaz}, \& {Jablonka}}]{Taibi2022}
{Taibi}, S., {Battaglia}, G., {Leaman}, R., {et~al.} 2022, \aap, 665, A92,
  \dodoi{10.1051/0004-6361/202243508}

\bibitem[{{Taibi} {et~al.}(2018){Taibi}, {Battaglia}, {Kacharov}, {Rejkuba},
  {Irwin}, {Leaman}, {Zoccali}, {Tolstoy}, \& {Jablonka}}]{Taibi2018}
{Taibi}, S., {Battaglia}, G., {Kacharov}, N., {et~al.} 2018, \aap, 618, A122,
  \dodoi{10.1051/0004-6361/201833414}

\bibitem[{{Teyssier} {et~al.}(2013){Teyssier}, {Pontzen}, {Dubois}, \&
  {Read}}]{Teyssier2013}
{Teyssier}, R., {Pontzen}, A., {Dubois}, Y., \& {Read}, J.~I. 2013, \mnras,
  429, 3068, \dodoi{10.1093/mnras/sts563}

\bibitem[{{Tollet} {et~al.}(2016){Tollet}, {Macci{\`o}}, {Dutton}, {Stinson},
  {Wang}, {Penzo}, {Gutcke}, {Buck}, {Kang}, {Brook}, {Di Cintio}, {Keller}, \&
  {Wadsley}}]{Tollet2016}
{Tollet}, E., {Macci{\`o}}, A.~V., {Dutton}, A.~A., {et~al.} 2016, \mnras, 456,
  3542, \dodoi{10.1093/mnras/stv2856}

\bibitem[{{Tolstoy} {et~al.}(2004){Tolstoy}, {Irwin}, {Helmi}, {Battaglia},
  {Jablonka}, {Hill}, {Venn}, {Shetrone}, {Letarte}, {Cole}, {Primas},
  {Francois}, {Arimoto}, {Sadakane}, {Kaufer}, {Szeifert}, \&
  {Abel}}]{Tolstoy2004}
{Tolstoy}, E., {Irwin}, M.~J., {Helmi}, A., {et~al.} 2004, \apjl, 617, L119,
  \dodoi{10.1086/427388}

\bibitem[{{Tortora} {et~al.}(2010){Tortora}, {Napolitano}, {Cardone},
  {Capaccioli}, {Jetzer}, \& {Molinaro}}]{Tortora2010}
{Tortora}, C., {Napolitano}, N.~R., {Cardone}, V.~F., {et~al.} 2010, \mnras,
  407, 144, \dodoi{10.1111/j.1365-2966.2010.16938.x}

\bibitem[{Tully {et~al.}(1996)Tully, Verheijen, Pierce, Huang, \&
  Wainscoat}]{Tully_1996}
Tully, R.~B., Verheijen, M. A.~W., Pierce, M.~J., Huang, J.-S., \& Wainscoat,
  R.~J. 1996, The Astronomical Journal, 112, 2471, \dodoi{10.1086/118196}

\bibitem[{{Wadsley} {et~al.}(2004){Wadsley}, {Stadel}, \&
  {Quinn}}]{Wadsley2004}
{Wadsley}, J.~W., {Stadel}, J., \& {Quinn}, T. 2004, \na, 9, 137,
  \dodoi{10.1016/j.newast.2003.08.004}

\bibitem[{{White} \& {Rees}(1978)}]{WhiteRees1978}
{White}, S.~D.~M., \& {Rees}, M.~J. 1978, \mnras, 183, 341,
  \dodoi{10.1093/mnras/183.3.341}

\bibitem[{{Williams} {et~al.}(2009){Williams}, {Dalcanton}, {Dolphin},
  {Holtzman}, \& {Sarajedini}}]{Williams2009}
{Williams}, B.~F., {Dalcanton}, J.~J., {Dolphin}, A.~E., {Holtzman}, J., \&
  {Sarajedini}, A. 2009, \apjl, 695, L15, \dodoi{10.1088/0004-637X/695/1/L15}

\bibitem[{{Wise} {et~al.}(2012){Wise}, {Abel}, {Turk}, {Norman}, \&
  {Smith}}]{Wise2012}
{Wise}, J.~H., {Abel}, T., {Turk}, M.~J., {Norman}, M.~L., \& {Smith}, B.~D.
  2012, \mnras, 427, 311, \dodoi{10.1111/j.1365-2966.2012.21809.x}

\bibitem[{{Zavala} {et~al.}(2016){Zavala}, {Frenk}, {Bower}, {Schaye},
  {Theuns}, {Crain}, {Trayford}, {Schaller}, \& {Furlong}}]{Zavala2016}
{Zavala}, J., {Frenk}, C.~S., {Bower}, R., {et~al.} 2016, \mnras, 460, 4466,
  \dodoi{10.1093/mnras/stw1286}

\bibitem[{{Zhang} {et~al.}(2012){Zhang}, {Hunter}, {Elmegreen}, {Gao}, \&
  {Schruba}}]{Zhang2012}
{Zhang}, H.-X., {Hunter}, D.~A., {Elmegreen}, B.~G., {Gao}, Y., \& {Schruba},
  A. 2012, \aj, 143, 47, \dodoi{10.1088/0004-6256/143/2/47}

\bibitem[{{Zheng} {et~al.}(2017){Zheng}, {Wang}, {Ge}, {Mao}, {Li}, {Li}, {Mo},
  {Goddard}, {Bundy}, {Li}, {Nair}, {Lin}, {Long}, {Riffel}, {Thomas},
  {Masters}, {Bizyaev}, {Brownstein}, {Zhang}, {Law}, {Drory}, {Roman Lopes},
  \& {Malanushenko}}]{Zheng2017}
{Zheng}, Z., {Wang}, H., {Ge}, J., {et~al.} 2017, \mnras, 465, 4572,
  \dodoi{10.1093/mnras/stw3030}

\bibitem[{{Zolotov} {et~al.}(2012){Zolotov}, {Brooks}, {Willman}, {Governato},
  {Pontzen}, {Christensen}, {Dekel}, {Quinn}, {Shen}, \&
  {Wadsley}}]{Zolotov2012}
{Zolotov}, A., {Brooks}, A.~M., {Willman}, B., {et~al.} 2012, \apj, 761, 71,
  \dodoi{10.1088/0004-637X/761/1/71}

\end{thebibliography}
\bibliographystyle{aasjournal}





\appendix

\section{Summary of Simulated Galaxy Properties}\label{AppA}

Here we present a table summarizing the main properties of our simulations. 

\startlongtable
\begin{deluxetable*}{lcccccccccl}
    \tabletypesize{\scriptsize}
    \tablewidth{0pt}
    \tablecaption{Properties of the simulated galaxies in our analysis}
    \tablehead{
\colhead{Halo} & \colhead{Mvir (M$_\odot$)} & \colhead{Mstar (M$_\odot$)} & \colhead{R$_e$ (kpc)} & \colhead{V-band Magnitude} & \colhead{t$_{90}$ (Gyr)} & \colhead{t$_{50}$ (Gyr)} & \colhead{dt$_{90}$/f$_{R_e}$ (Gyr)} & \colhead{dt$_{50}$/f$_{R_e}$ (Gyr)} & \colhead{Environment}
}
\startdata
Cpt. Marvel 1 & 1.55e+10 & 3.26e+07 & 1.2 & -13.8 & 10.7 & 2.0 & $-2.4^{+1.0}_{-0.5}$ & $-2.8^{+0.7}_{-1.1}$ & field\\
Cpt. Marvel 2 & 9.88e+09 & 8.95e+06 & 1.9 & -13.0 & 12.7 & 6.9 & $-0.7^{+0.2}_{-0.7}$ & $-2.9^{+0.2}_{-1.2}$ & field\\
Cpt. Marvel 3 & 8.75e+09 & 5.11e+06 & 0.8 & -12.5 & 12.9 & 7.6 & $-1.2^{+0.1}_{-0.2}$ & $-3.2^{+0.1}_{-0.3}$ & field\\
Cpt. Marvel 5 & 7.50e+09 & 7.33e+06 & 0.6 & -12.1 & 8.7 & 2.2 & $-4.0^{+0.4}_{-0.8}$ & $-0.9^{+0.1}_{-0.2}$ & field\\
Cpt. Marvel 6 & 6.58e+09 & 6.88e+06 & 0.5 & -11.9 & 7.5 & 1.9 & $-6.1^{+0.8}_{-0.6}$ & $-2.4^{+0.2}_{-0.1}$ & field\\
Cpt. Marvel 7 & 5.10e+09 & 6.85e+05 & 0.3 & -10.4 & 13.2 & 2.4 & $-0.3^{+0.1}_{-0.3}$ & $-2.1^{+0.5}_{-5.9}$ & field\\
Cpt. Marvel 10 & 3.48e+09 & 2.06e+06 & 0.4 & -10.1 & 2.2 & 1.6 & $-0.1^{+0.1}_{-0.1}$ & $-0.2^{+0.0}_{-0.0}$ & field\\
Cpt. Marvel 13 & 1.80e+09 & 1.83e+05 & 0.3 & -7.7 & 2.4 & 2.3 & $0.0^{+0.0}_{-0.0}$ & $0.2^{+0.1}_{-0.2}$ & field\\
Elektra 1 & 4.18e+10 & 1.49e+08 & 2.2 & -16.0 & 12.8 & 7.2 & $-1.1^{+0.4}_{-0.6}$ & $-4.2^{+1.1}_{-1.4}$ & field\\
Elektra 2 & 2.90e+10 & 4.39e+07 & 0.7 & -15.0 & 13.3 & 5.5 & $-0.5^{+0.2}_{-0.2}$ & $-4.3^{+0.1}_{-0.2}$ & field\\
Elektra 3 & 2.45e+10 & 2.54e+07 & 1.5 & -14.3 & 12.9 & 8.2 & $-0.8^{+0.1}_{-0.5}$ & $-1.6^{+0.3}_{-1.0}$ & field\\
Elektra 4 & 1.91e+10 & 3.65e+07 & 1.0 & -13.8 & 11.0 & 2.1 & $-1.4^{+0.3}_{-0.2}$ & $1.3^{+0.2}_{-0.4}$ & field\\
Elektra 5 & 1.53e+10 & 2.35e+07 & 0.7 & -13.9 & 12.3 & 3.0 & $-0.8^{+0.6}_{-1.0}$ & $-6.0^{+0.1}_{-0.2}$ & field\\
Elektra 10 & 4.11e+09 & 7.05e+05 & 0.3 & -10.0 & 12.1 & 2.9 & $-1.3^{+0.2}_{-0.6}$ & $-5.5^{+0.4}_{-1.8}$ & field\\
Storm 1 & 7.62e+10 & 4.72e+08 & 4.6 & -17.6 & 13.3 & 8.9 & $1.6^{+0.2}_{-0.6}$ & $2.2^{+1.0}_{-3.4}$ & field\\
Storm 2 & 3.52e+10 & 9.48e+07 & 3.2 & -15.6 & 12.9 & 8.6 & $-0.5^{+0.2}_{-0.2}$ & $0.7^{+0.3}_{-0.3}$ & field\\
Storm 3 & 2.43e+10 & 2.34e+07 & 1.1 & -14.3 & 13.1 & 3.6 & $-1.2^{+0.3}_{-0.4}$ & $-10.3^{+1.6}_{-2.0}$ & field\\
Storm 4 & 1.68e+10 & 3.61e+07 & 1.8 & -14.1 & 11.4 & 8.0 & $-2.0^{+0.2}_{-0.3}$ & $-0.7^{+0.1}_{-0.3}$ & field\\
Storm 5 & 9.62e+09 & 7.95e+06 & 0.4 & -12.5 & 11.5 & 2.6 & $-1.8^{+0.2}_{-0.4}$ & $-7.4^{+0.3}_{-0.6}$ & field\\
Storm 6 & 8.43e+09 & 3.33e+06 & 0.3 & -12.3 & 13.3 & 8.8 & $-0.7^{+0.3}_{-0.0}$ & $-3.7^{+0.1}_{-0.8}$ & field\\
Storm 7 & 8.02e+09 & 9.79e+06 & 0.6 & -12.2 & 5.0 & 1.8 & $-6.3^{+1.9}_{-1.6}$ & $-0.5^{+0.0}_{-0.2}$ & field\\
Storm 8 & 7.95e+09 & 1.04e+07 & 0.7 & -12.8 & 11.9 & 2.7 & $-2.1^{+0.2}_{-0.5}$ & $-6.3^{+0.7}_{-0.7}$ & field\\
Storm 12 & 5.65e+09 & 1.60e+06 & 0.3 & -10.9 & 11.7 & 2.8 & $-1.2^{+0.2}_{-0.6}$ & $-3.9^{+0.2}_{-0.4}$ & field\\
Storm 14 & 3.66e+09 & 8.83e+05 & 0.3 & -9.5 & 2.5 & 1.4 & $-13.7^{+2.3}_{-3.8}$ & $-0.6^{+0.1}_{-0.3}$ & field\\
Storm 23 & 1.84e+09 & 2.30e+05 & 0.3 & -7.9 & 2.3 & 1.8 & $0.1^{+0.1}_{-0.1}$ & $-0.0^{+0.0}_{-0.0}$ & field\\
Storm 31 & 9.95e+08 & 3.02e+06 & 0.5 & -11.3 & 10.1 & 2.9 & $-3.1^{+0.2}_{-0.3}$ & $-4.1^{+0.4}_{-0.9}$ & satellite\\
Rogue 1 & 8.32e+10 & 9.86e+08 & 0.7 & -17.8 & 13.1 & 7.3 & $-0.4^{+0.0}_{-0.0}$ & $-1.5^{+0.4}_{-0.4}$ & field\\
Rogue 3 & 2.12e+10 & 6.96e+07 & 3.4 & -14.9 & 11.5 & 6.5 & $0.8^{+0.4}_{-0.5}$ & $-0.0^{+0.0}_{-0.0}$ & field\\
Rogue 7 & 1.47e+10 & 2.34e+07 & 0.9 & -13.7 & 11.8 & 3.9 & $-2.1^{+0.7}_{-0.2}$ & $-5.2^{+1.3}_{-0.1}$ & field\\
Rogue 8 & 1.30e+10 & 5.11e+07 & 1.6 & -14.2 & 9.7 & 2.0 & $-3.1^{+0.2}_{-0.3}$ & $-0.8^{+0.1}_{-0.3}$ & field\\
Rogue 10 & 1.13e+10 & 1.40e+07 & 0.7 & -12.7 & 8.1 & 3.9 & $-6.3^{+1.7}_{-0.2}$ & $-4.6^{+0.6}_{-1.0}$ & field\\
Rogue 11 & 8.09e+09 & 4.67e+06 & 1.0 & -12.2 & 12.5 & 6.9 & $0.3^{+0.2}_{-0.2}$ & $-1.7^{+0.3}_{-0.5}$ & field\\
Rogue 12 & 7.61e+09 & 8.36e+06 & 0.8 & -12.3 & 10.3 & 2.6 & $-2.8^{+0.3}_{-1.6}$ & $-6.7^{+0.5}_{-1.3}$ & field\\
Rogue 28 & 1.64e+09 & 9.05e+05 & 0.5 & -9.4 & 2.6 & 1.7 & $-0.3^{+0.1}_{-0.1}$ & $-0.0^{+0.0}_{-0.0}$ & field\\
Rogue 31 & 1.56e+09 & 2.56e+05 & 0.3 & -8.0 & 1.8 & 1.1 & $0.1^{+0.0}_{-0.0}$ & $-0.0^{+0.0}_{-0.1}$ & field\\
Rogue 37 & 1.10e+09 & 1.00e+06 & 0.4 & -10.0 & 11.9 & 2.2 & $-1.1^{+0.2}_{-1.2}$ & $-4.6^{+1.1}_{-4.9}$ & satellite\\
Sandra 2 & 9.58e+10 & 1.36e+09 & 3.0 & -18.7 & 13.3 & 9.3 & $-0.8^{+0.1}_{-0.2}$ & $-3.9^{+0.4}_{-1.3}$ & satellite\\
Sandra 3 & 4.70e+10 & 9.01e+08 & 4.0 & -17.6 & 12.3 & 5.6 & $1.2^{+0.1}_{-0.5}$ & $-1.1^{+0.8}_{-0.9}$ & satellite\\
Sandra 4 & 3.25e+10 & 2.59e+08 & 2.5 & -16.2 & 12.1 & 4.3 & $-1.2^{+0.5}_{-1.3}$ & $-1.8^{+1.1}_{-2.6}$ & satellite\\
Sandra 6 & 2.87e+10 & 1.99e+08 & 2.4 & -16.1 & 12.4 & 4.4 & $-1.6^{+0.4}_{-0.4}$ & $-4.5^{+0.6}_{-1.3}$ & field\\
Sandra 7 & 1.56e+10 & 1.42e+08 & 1.2 & -15.1 & 9.3 & 2.9 & $-3.2^{+0.0}_{-0.2}$ & $-4.0^{+0.3}_{-0.6}$ & field\\
Sandra 11 & 1.20e+10 & 1.47e+07 & 0.8 & -13.0 & 10.8 & 4.8 & $-2.2^{+1.3}_{-1.6}$ & $-1.7^{+1.1}_{-0.1}$ & field\\
Sandra 12 & 9.07e+09 & 6.26e+07 & 0.8 & -14.0 & 4.5 & 1.7 & $-5.0^{+0.2}_{-0.9}$ & $-0.9^{+0.1}_{-0.4}$ & satellite\\
Sandra 13 & 8.57e+09 & 7.45e+06 & 0.7 & -12.6 & 12.1 & 7.9 & $-1.8^{+0.1}_{-0.1}$ & $-2.1^{+0.1}_{-1.1}$ & satellite\\
Sandra 15 & 6.45e+09 & 2.46e+06 & 0.6 & -11.3 & 11.9 & 8.3 & $-0.5^{+1.3}_{-1.4}$ & $-0.9^{+0.6}_{-1.9}$ & field\\
Sandra 20 & 4.55e+09 & 2.34e+06 & 0.3 & -11.4 & 9.3 & 3.1 & $-5.7^{+1.3}_{-3.5}$ & $-3.7^{+1.0}_{-5.9}$ & field\\
Sandra 23 & 3.33e+09 & 6.02e+07 & 2.2 & -14.5 & 10.8 & 6.9 & $-2.5^{+0.1}_{-0.5}$ & $-1.2^{+0.0}_{-0.4}$ & satellite\\
Sandra 27 & 3.25e+09 & 5.05e+07 & 1.7 & -14.0 & 8.9 & 4.3 & $-3.9^{+1.6}_{-0.9}$ & $-2.7^{+0.9}_{-0.6}$ & field\\
Sandra 28 & 3.44e+09 & 1.90e+06 & 0.5 & -10.9 & 11.8 & 5.4 & $-4.2^{+1.3}_{-2.8}$ & $-6.2^{+2.7}_{-7.0}$ & satellite\\
Sandra 33 & 2.68e+09 & 2.22e+06 & 0.6 & -10.4 & 5.3 & 1.6 & $-4.2^{+2.7}_{-2.8}$ & $-1.0^{+0.3}_{-0.9}$ & field\\
Sandra 34 & 2.63e+09 & 3.46e+06 & 1.3 & -11.3 & 10.1 & 6.1 & $-2.6^{+0.3}_{-0.9}$ & $-3.3^{+0.6}_{-2.9}$ & field\\
Sandra 38 & 1.88e+09 & 8.24e+06 & 1.5 & -11.9 & 7.7 & 2.8 & $-2.2^{+0.4}_{-0.5}$ & $-1.9^{+0.1}_{-0.9}$ & satellite\\
Sandra 65 & 9.00e+08 & 1.80e+07 & 2.4 & -12.6 & 4.4 & 2.3 & $-1.5^{+0.1}_{-0.3}$ & $-0.8^{+0.1}_{-0.2}$ & satellite\\
Sandra 114 & 5.11e+08 & 7.85e+05 & 0.5 & -9.2 & 3.2 & 1.8 & $0.2^{+0.9}_{-1.6}$ & $-0.0^{+0.5}_{-0.3}$ & field\\
Ruth 2 & 5.11e+10 & 5.00e+08 & 1.7 & -17.1 & 12.6 & 4.6 & $-0.9^{+0.5}_{-1.3}$ & $-4.6^{+0.4}_{-0.8}$ & field\\
Ruth 3 & 1.10e+10 & 2.09e+07 & 0.9 & -13.0 & 6.4 & 2.5 & $-6.8^{+0.2}_{-2.5}$ & $-1.7^{+0.1}_{-0.6}$ & field\\
Ruth 6 & 8.29e+09 & 1.49e+07 & 1.5 & -13.1 & 11.2 & 4.1 & $-2.0^{+0.3}_{-1.0}$ & $-1.6^{+0.7}_{-0.9}$ & field\\
Ruth 14 & 2.40e+09 & 2.00e+06 & 0.5 & -10.7 & 10.3 & 6.4 & $-3.0^{+0.6}_{-1.1}$ & $-1.5^{+1.6}_{-2.3}$ & field\\
Ruth 18 & 1.51e+09 & 4.63e+07 & 1.6 & -13.6 & 5.1 & 2.3 & $-4.4^{+0.5}_{-0.8}$ & $-1.0^{+0.1}_{-0.2}$ & satellite\\
Ruth 22 & 1.55e+09 & 2.87e+06 & 0.8 & -10.6 & 3.7 & 1.9 & $-2.2^{+1.6}_{-2.6}$ & $-0.9^{+0.2}_{-0.4}$ & satellite\\
Ruth 49 & 7.64e+08 & 4.07e+06 & 0.9 & -11.1 & 6.5 & 2.1 & $-3.8^{+0.5}_{-0.7}$ & $-2.3^{+0.3}_{-0.7}$ & satellite\\
Ruth 92 & 3.50e+08 & 1.38e+06 & 0.6 & -9.8 & 3.1 & 1.4 & $0.0^{+0.0}_{-1.9}$ & $-0.1^{+0.1}_{-0.4}$ & satellite\\
Sonia 8 & 1.19e+10 & 4.80e+08 & 2.2 & -16.7 & 11.1 & 4.7 & $-2.2^{+0.3}_{-0.5}$ & $-3.0^{+0.7}_{-1.2}$ & satellite\\
Sonia 10 & 9.26e+09 & 4.77e+08 & 1.7 & -16.4 & 9.5 & 4.6 & $-1.8^{+0.4}_{-1.1}$ & $-1.3^{+0.1}_{-0.1}$ & satellite\\
Sonia 21 & 4.01e+09 & 8.20e+06 & 0.7 & -11.8 & 7.9 & 2.5 & $-1.8^{+0.1}_{-0.2}$ & $-1.4^{+0.3}_{-1.2}$ & satellite\\
Sonia 30 & 2.77e+09 & 2.32e+06 & 0.6 & -9.9 & 5.3 & 2.7 & $-0.2^{+0.2}_{-1.8}$ & $-1.1^{+1.4}_{-3.7}$ & field\\
Sonia 34 & 2.11e+09 & 6.46e+05 & 0.4 & -9.0 & 1.8 & 0.5 & $0.9^{+0.4}_{-1.4}$ & $-0.0^{+0.0}_{-0.4}$ & field\\
Sonia 38 & 1.98e+09 & 1.10e+06 & 0.3 & -10.0 & 10.9 & 3.2 & $-1.4^{+0.5}_{-1.1}$ & $-5.1^{+2.4}_{-4.0}$ & field\\
Elena 7 & 5.13e+09 & 1.65e+07 & 1.0 & -13.1 & 10.5 & 2.7 & $-1.0^{+0.4}_{-1.1}$ & $-1.4^{+2.0}_{-0.8}$ & satellite\\
Elena 29 & 1.40e+09 & 8.95e+05 & 0.5 & -9.6 & 7.5 & 3.5 & $-0.9^{+0.6}_{-2.9}$ & $-2.1^{+0.6}_{-1.6}$ & field\\
Elena 115 & 2.61e+08 & 6.58e+05 & 0.9 & -9.1 & 3.6 & 2.4 & $-0.4^{+0.4}_{-0.6}$ & $-0.0^{+0.4}_{-0.8}$ & satellite\\
Elena 117 & 2.40e+08 & 1.85e+06 & 0.6 & -10.1 & 3.5 & 1.3 & $0.4^{+0.3}_{-0.5}$ & $-0.4^{+0.9}_{-0.5}$ & satellite\\
\enddata
\label{Table1}
\end{deluxetable*}

\section{The effect of dark matter core creation on age gradient formation}\label{AppB}
\begin{figure*}
    \centering
    \includegraphics[width=0.85\textwidth]{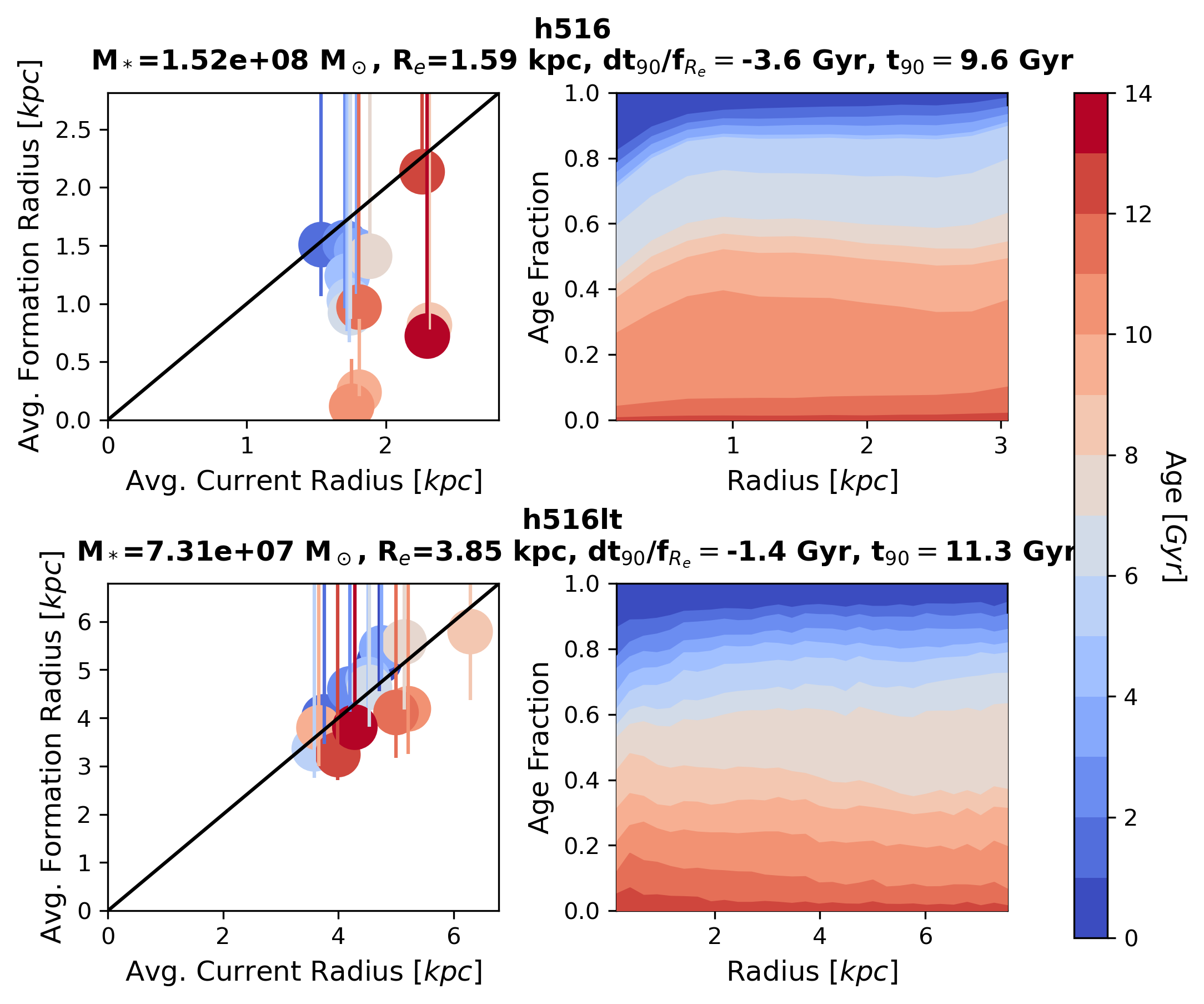}
    \caption{Here, we examine two galaxies run with different star formation physics: h516 and h516lt.  h516 is run with a high star formation density threshold (requiring the presence of H$_2$) and forms a dark matter core, while h516lt has a low density threshold for star formation and does not form a core.   We remake the left and middle panels from Fig.~\ref{fig:r_age_plots} for these two galaxies. We also label the stellar mass, effective radius, age gradient, and global $t_{90}$ value. We find that only h516 has significant reshuffling of its oldest stars, consistent with the idea that the same process that creates dark matter cores also reshuffles the stars.}
    \label{fig:h516}
\end{figure*}

In order to determine if core creation does indeed lead to outside-in formation, we examine two versions of the same simulated galaxy, named h516 and h516lt. These two simulations were run with ChaNGa's predecessor code, Gasoline.  They use a WMAP3 cosmology, and are zoom simulations run within a (25 Mpc)$^3$ volume.  They have nearly identical SN feedback to the simulations in this work, i.e., the blastwave feedback model, but inject 1.0$\times$10$^{51}$ erg of thermal energy per SN instead of 1.5$\times$10$^{51}$ erg.  While the star formation physics for h516 is identical to the star formation used through the rest of this work, h516lt uses a different star formation prescription that has been shown to prevent dark matter core formation.  In particular, it has been found that star formation must occur above a certain density threshold in order to create dark matter cores in simulations \citep[e.g.,][]{Governato2010, Dutton2019}.  Thus, h516lt instead allows stars to form when gas particles have $T < 10^4$ K and $\rho < 0.1$ cm$^{-3}$.  h516 has been shown \cite[e.g.,][]{Governato2012} to create a dark matter core, with $\alpha \sim -0.2$.  We verify that h516lt maintains a cuspy dark matter density profile, with $\alpha = -1.1$.  Thus, the two versions of h516 provide us with a case-study to further examine how dark matter core creation is related to stellar age gradients of dwarfs.  


We recreate the left and middle panels of Fig.~\ref{fig:r_age_plots} for both versions of the galaxy and show the results in Fig.~\ref{fig:h516}.  The global $t_{90}$ and $dt_{90}/f_{R_e}$ values for each galaxy are listed in the plot headers.  h516 a steeper, outside-in age gradient ($-3.6$ Gyr) as well as an earlier $t_{90}$ value (9.6 Gyr) than h516lt.  These results are consistent with h516 lying on the same relation for the simulated galaxies in Figure~\ref{fig:dt90_t90_overview}.  On the other hand, it can be seen that h516lt has a flatter gradient.  With $t_{90} = 11.3$ Gyr and $t_{90}$ and $dt_{90}/f_{R_e} = -1.4$ Gyr, h516lt would lie slightly above the majority of galaxies in Figure~\ref{fig:dt90_t90_overview}, though still be consistent within errors.  However, we note that h516lt would lie significantly off the mean size-mass relation for dwarf galaxies \citep[e.g.,][]{Brooks2011}.  The high star formation threshold in h516, requiring the presence of H$_2$, limits star formation to be more centralized, where densities are higher, while the lower threshold in h516lt allows star formation to unrealistic radii. 

The left panels of Fig.~\ref{fig:h516} make it clear that only h516,  the galaxy with a high enough star formation threshold to form a dark matter core, has had its old stars significantly reshuffled.  
In h516lt, both the oldest and youngest stars form at roughly the same radius (on average) and do not move substantially relative to where they are formed. We conclude that stellar feedback in the clustered, multi-phase ISM drives both dark matter core creation and the stellar reshuffling which produces steep age gradients.




\end{document}